\documentclass[reprint,nofootinbib,showpacs]{revtex4}
\usepackage{graphicx}  
\usepackage{dcolumn}   
\usepackage{bm}        
\usepackage{amssymb}
\usepackage{hyperref}

\begin{document}
\title{Implication of $R_{D^{(*)}}$ anomalies on semileptonic decays of $\Sigma_b$ and $\Omega_b$ baryons}
\author{N~Rajeev${}^{1}$}
\email{rajeev@rs.phy.student.nits.ac.in} 
\author{Rupak~Dutta${}^{1}$}
\email{rupak@phy.nits.ac.in}
\affiliation{
${}^1$National Institute of Technology Silchar, Silchar 788010, India\\
}
\author{Suman Kumbhakar${}^{2}$}
\email{suman@phy.iitb.ac.in}
\affiliation{${}^2$Indian Institute of Technology Bombay, Mumbai 400076, India
}

\begin{abstract}
The flavor changing decays of heavy bottom quarks to the corresponding lighter quarks ($u$, $c$ and $s$) in various $B$-meson decays via charged current
and neutral current semileptonic transitions have emerged as promising candidates to explore the physics beyond the standard model.  
Experimentally the lepton flavor universality violation in $b \to (c,u)\,l\,\nu$ and $b \to s\,l^+ l^-$ transitions 
have been reported to a higher precision. 
The measurements of the lepton flavor violating
ratios such as $R_{D^{(*)}}$, $R_{J/\Psi}$ and $R_{K^{(*)}}$ are observed to deviate from the standard model expectations at 
the level of $1.4\sigma$, $2.5\sigma$, $1.5\sigma$, $2.4\sigma$ and $2.2\sigma$ respectively. 
Motivated by these anomalies, we investigate the lepton flavor universality violation in $\Sigma_b \to \Sigma_c l \nu$ and
$\Omega_b \to \Omega_c l \nu$ decays. We follow a model independent effective field theory formalism 
and study the implications of $R_{D^{(*)}}$ anomalies on
$\Sigma_b \to \Sigma_c \tau \nu$ and $\Omega_b \to \Omega_c \tau \nu$ decay modes.
We give predictions of various physical observables such as the ratio of branching ratios,
total differential decay rate, forward-backward asymmetry, lepton side polarization fraction and convexity parameter within the standard model and within
various new physics scenarios.
\end{abstract}


\maketitle

\section{Introduction}
\begin{figure}[htbp]
\centering
\includegraphics[width=9.4cm,height=4.5cm]{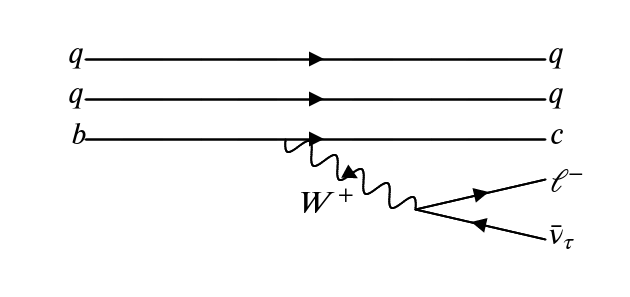}
\caption{A tree level Feynman diagram representing the transitions of $\Sigma_b^- \,(ddb) \to \Sigma_c^0 \,(ddc) \, l^- \, \bar{\nu}_l$ and
$\Omega_b^- \,(ssb) \to \Omega_c^0 \,(ssc) \, l^- \, \bar{\nu}_l$.}
\label{feyn}
\end{figure}

Although, the present day experimental results in $B$ factory experiments are dominated by the meson decays over the baryon decays, 
the theoretical exploration of the semileptonic decays of baryons have a longer history as compared to the mesons.
The system of particles which are classified under mesons and baryons are mainly distinguished by their quark structure. In early 1960's 
the concept of diquarks emerged out of some critical phenomenological ideas, have lead to the diverse coherent thoughts about the baryon
decay characteristics. Soon after in Ref.~\cite{Ida:1966ev,Lichtenberg:1967zz} the 
concept of diquark was literally introduced in order to describe a baryon as a composite state of two particles called a quark and a diquark.
The heavy quark symmetry assumes baryons as a bound state of $(Q\,q\,q)$ where, $Q$ being the heavy quark surrounded by the lighter
quarks $q$. This idea of quark-diquark picture of a baryon have successfully managed to predict various properties including their compositions
and the decay probabilities.
During the weak decays of baryons, only the heavier quark will be knocked out of a baryon and take part in the decay process by changing its flavor
whereas, the lighter diquark pair will act as a spectator~\cite{Anselmino:1992vg}. This is because when we carefully monitor this process, the quantum numbers (color index,
helicities, momentum) are conserved for lighter diquark system. Hence, consequently this baryon three-body problem is reduced to usual meson
two-body problem (see Fig.~\ref{feyn}).
Therefore, at the scale of quark level transitions, the treatment of semileptonic decays of baryons are considered to be very much analogous to that 
of mesons.

Study of semileptonic $B$ meson decays is of great interest due to the long standing anomalies that are present in various $B$ meson decays
mediated via $b \to c\,l\, \nu$ and $b \to s\,l^+ l^-$ quark level transitions.
The most well grounded measurements which substantiate these anomalies are the ratio of branching ratios $R_D$ and $R_{D^{*}}$ defined as,
\begin{equation}
 R_D = \frac{{\mathcal{B}(B\to D \tau \nu)}}{{\mathcal{B}(B\to D \{e/\mu\} \nu)}}, \hspace{1cm}
 R_{D^*} = \frac{{\mathcal{B}(B\to D^* \tau \nu)}}{{\mathcal{B}(B\to D^* \{e/\mu\} \nu)}}.
\end{equation}

The precise SM predictions of $R_D$ and $R_{D^{*}}$ based on the recent lattice calculations have been carried out by various groups and 
interestingly every predictions are in good agreement with each other. The FNAL/MILC Collaboration predicts the value of $R_D$ to be 
$0.299 \pm 0.011$~\cite{Lattice:2015rga}. Similarly, in Ref.~\cite{Na:2015kha} it is predicted to be $0.300 \pm 0.008$. By combining these two calculations
the FLAG working group~\cite{Aoki:2016frl} have come up with a value of $R_D = 0.300 \pm 0.008$. The authors in Ref.~\cite{Bigi:2016mdz} suggest more accurate value
of $R_D = 0.299 \pm 0.003$ by combing the two lattice calculations by obtaining the experimental form factors of $B \to D l \nu$ from $BABAR$ 
and Belle. In fact various similar calculations of $R_D$ can also be found in Refs.~\cite{Bernlochner:2017jka,Jaiswal:2017rve}. 
Regarding the $R_{D^{*}}$ SM predictions, at present we have quite a large number of predictions in which every prediction manifests
a minimal variation. In Ref.~\cite{Fajfer:2012vx} the authors predicts the value to be $R_{D^{*}} = 0.252 \pm 0.003$. More recent calculations of $R_{D^{*}}=$
$0.257 \pm 0.003$~\cite{Bernlochner:2017jka}, $0.257 \pm 0.005$~\cite{Jaiswal:2017rve} and $0.260 \pm 0.008$~\cite{Bigi:2017jbd} obtained 
from the new form factor inputs by fitting the unfolded spectrum from Belle with the BGL parametrization~\cite{Abdesselam:2017kjf} are in good agreement 
with each other as well as with the previous prediction. One can expect even more precise prediction of $R_{D^{*}}$ once the full lattice QCD calculations
are available.
On the other hand, we have several measurements of $R_D$ and $R_{D^{*}}$ from various experiments such as BABAR, Belle and LHCb.
The Heavy Flavor Averaging Group~(HFLAV) determined the combined deviation in $R_{D^{(*)}}$ with respect to the SM. 
Recent measurements from Belle in 2019 have a significant impact on the average values of $R_{D^{(*)}}$.
At present the combined deviation in $R_{D^{(*)}}$ is reported to be $3.08\sigma$ from the SM expectations. 
The average values of $R_D$ and $R_{D^{*}}$ reported by HFLAV are displayed in Table~\ref{inputs3}.

\begin{table}[ht!]
\centering
{\begin{tabular}{|c|c|c|c|}
\hline
Observables & SM predictions & World averages & Deviation\\
\hline
\hline
 $R_D = {\mathcal{B}(B\to D \tau \nu)}/{\mathcal{B}(B\to D l \nu)}$ & $0.299 \pm 0.003$~\cite{Lattice:2015rga,Na:2015kha,Aoki:2016frl,Bigi:2016mdz} & $0.340 \pm 0.027 \pm 0.013$~\cite{Lees:2013uzd,Lees:2012xj,Huschle:2015rga,Abdesselam:2019dgh} & $1.4\sigma$\\

 $R_{D^*} = {\mathcal{B}(B\to D^* \tau \nu)}/{\mathcal{B}(B\to D^* l \nu)}$ & $0.258 \pm 0.005$~\cite{Fajfer:2012vx,Bernlochner:2017jka,Jaiswal:2017rve,Bigi:2017jbd} & $0.295 \pm 0.011 \pm 0.008$~\cite{Lees:2013uzd,Lees:2012xj,Huschle:2015rga,Abdesselam:2019dgh,Hirose:2016wfn,Sato:2016svk,Hirose:2017dxl,Aaij:2015yra,Aaij:2017uff,Aaij:2017deq} & $2.5\sigma$\\

\hline 
\hline
\end{tabular}}
\caption{Recent SM predictions and world averages of $R_D$ and $R_{D^{*}}$.}
\label{inputs3}
\end{table}

The clear disagreements between the SM predictions and the experimental measurements strongly indicate towards possible new physics.
Several new physics scenarios are being investigated within model dependent and model independent frameworks~\cite{Sakaki:2014sea,
Freytsis:2015qca,Bhattacharya:2016zcw,Alok:2017qsi,Azatov:2018knx,Bifani:2018zmi,Huang:2018nnq,Hu:2018veh,Feruglio:2018fxo,Iguro:2018fni,
Jung:2018lfu,Datta:2017aue,Dutta:2018jxz,Bernlochner:2018kxh,Dutta:2017xmj,Alok:2018uft,Dutta:2018vgu}. 
Similarly, implications of $R_{D^{(*)}}$ anomalies on similar decay modes have been studied as well. 
The details can be found in Refs.~\cite{Fajfer:2012jt,Crivellin:2012ye,Li:2016vvp,Bardhan:2016uhr,Alok:2016qyh,Dutta:2016eml,Dutta:2017wpq,
Bhattacharya:2016mcc,Dutta:2015ueb,Rajeev:2018txm,Alok:2019uqc}.

Apart from $R_D$ and $R_{D^{*}}$ measurements, the LHCb have also measured the ratio of branching ratio $R_{J/\Psi} = 
{\mathcal{B}(B_c\to J/\Psi \tau \nu)}/{\mathcal{B}(B_c\to J/\Psi l \nu)}$ to be $0.71 \pm 0.17 \pm 0.18$~\cite{Aaij:2017tyk}
which stands around $1.3\sigma$ away from the SM value of $[0.20,\,0.39]$~\cite{Cohen:2018dgz}. 
As this error is relatively large, we do not consider $R_{J/\Psi}$ in our new physics (NP) analysis.

In the SM, the $\Sigma_b$ and $\Omega_b$ semileptonic decays have been studied by several authors using the $\Sigma_b \to \Sigma_c$ and 
$\Omega_b \to \Omega_c$ transition form factors obtained in spectator-quark model, the relativistic quark model, the Bethe-Salpeter 
approach, relativistic three-quark model, the light-front quark model~\cite{Ebert:2006rp,Singleton:1990ye,Ivanov:1996fj,Ivanov:1998ya,
Ke:2012wa,Wang:2017mqp,
Korner:1992uw,Isgur:1990pm,Georgi:1990cx,Cheng:1995fe,Korner:1994nh,Ke:2019smy}. The total decay rate $\Gamma$ (in units of $10^{10}$s$^{-1}$) 
predicted within these models ranges from 1.44 - 2.23 for $\Sigma_b \to \Sigma_c e \nu$ and 1.29 - 1.87 for $\Omega_b \to \Omega_c e \nu$.
These variations in the prediction of $\Gamma$ may due to the complexity in understanding the baryon structures 
and also due to the lack of precise predictions of various form factors.
Nevertheless, we explore the NP effects on various observables pertaining to $\Sigma_b \to \Sigma_c \tau \nu$ and 
$\Omega_b \to \Omega_c \tau \nu$ decays within the model independent effective field theory formalism. 
It is indeed essential to study these decay modes both theoretically and experimentally to test the lepton flavor universality 
violation~(LFUV). 

Investigating the implications of $R_{D^{(*)}}$ on $\Sigma_b \to \Sigma_c \tau \nu$ and 
$\Omega_b \to \Omega_c \tau \nu$ decays will draw more interesting results. 
For this study, we have considered the form factors obtained in the relativistic quark model~\cite{Ebert:2006rp}. We give predictions of
various observables within the SM and within various NP scenarios. 
The results pertaining to the lepton side forward-backward asymmetry and the convexity parameter are predicted in SM for the first time in 
both the decay modes. 
Also, the new physics studies on these particular decay modes have not been explored till today. 
 
The paper is organized as follows: in Sec.~\ref{meth}, we briefly review the effective Lagrangian in the presence of the new physics 
couplings. 
Next we discuss the helicity formalism for $\Sigma_b \to \Sigma_c$ and $\Omega_b \to \Omega_c$ transitions and write down the respective
vector, axial-vector, scalar and pseudoscalar helicity amplitudes. We also write down the formulae to calculate the total differential 
decay rate
and various $q^2$ dependant observables. In Sec.~\ref{results}, we discuss the numerical results with all necessary input parameters. The 
numerical results are reported within the SM and within various NP scenarios. Finally, 
we conclude with a brief summary of our results in Sec.~\ref{summary}.

\section{Methodology}
\label{meth} 
Effective field theory formalism is a natural way to separate the effects coming from different scales involved in weak decays.
The most relevant effective Hamiltonian for $b \to c\,l\,\nu$ transition decays represented at the scale of bottom quark, containing both the SM and 
the possible NP operators is defined as~\cite{Cirigliano:2009wk,Bhattacharya:2011qm},
\begin{eqnarray}
\mathcal{H}_{eff} &=& \frac{4G_F}{\sqrt{2}} V_{cb} \Bigg[ \left( 1 + {V_L} \right) \mathcal{O}_{V_L} + {V_R} \mathcal{O}_{V_R} +
{S_L} \mathcal{O}_{S_L} + {S_R} \mathcal{O}_{S_R} + {T} \mathcal{O}_{T} \Bigg]+ {\rm h.c.} \,,
\end{eqnarray}
where $G_F$ is the Fermi coupling constant, $V_{cb}$ is the CKM matrix element, 
${V_L}$, ${V_R}$, ${S_L}$, ${S_R}$ and ${T}$ are the Wilson coefficients (WCs) corresponding to
the vector, scalar and tensor NP operators. The Fermionic operators $\mathcal{O}_{V_L}$, $\mathcal{O}_{V_R}$, $\mathcal{O}_{S_L}$, 
$\mathcal{O}_{S_R}$ and $\mathcal{O}_{T}$ are defined as,
\begin{equation}
 \mathcal{O}_{V_L} = \left( \bar{c} \gamma^{\mu} b_L \right) \left(\bar{l}_L \gamma_{\mu} \nu_{l\,L} \right)\,, \qquad\qquad
 \mathcal{O}_{V_R} = \left( \bar{c} \gamma_{\mu} b_R \right) \left(\bar{l}_L \gamma_{\mu} \nu_{l\,L} \right)
\end{equation}
\begin{equation}
 \mathcal{O}_{S_L} = \left( \bar{c} b_L \right) \left(\bar{l}_R \nu_{l\,L} \right)\,, \qquad\qquad
 \mathcal{O}_{S_R} = \left( \bar{c} b_R \right) \left(\bar{l}_R \nu_{l\,L} \right)
\end{equation}
\begin{equation}
 \mathcal{O}_{T} = \left( \bar{c} \sigma^{\mu \nu} b_L \right) \left(\bar{l}_R \sigma_{\mu \nu} \nu_{l\,L} \right).
\end{equation}

Here, we assume the neutrino to be always left chiral and all the WCs to be real.
We rewrite the effective Lagrangian by considering NP contributions only from the vector and scalar type interactions as~\cite{Dutta:2013qaa}
\begin{eqnarray}
\label{effl}
\mathcal{L}_{eff} &=&-\frac{G_F}{\sqrt{2}}\,V_{cb}\,\Bigg\{G_V\,\bar{l}\,\gamma_{\mu}\,(1 - \gamma_5)\,\nu_l\,\bar{c}\,\gamma^{\mu}\,b -
G_A\,\bar{l}\,\gamma_{\mu}\,(1 - \gamma_5)\,\nu_l\,\bar{c}\,\gamma^{\mu}\,\gamma_5\,b + \nonumber \\
&&  
G_S\,\bar{l}\,\gamma_{\mu}\,(1 - \gamma_5)\,\nu_l\,\bar{c}\,b -
G_P\,\bar{l}\,\gamma_{\mu}\,(1 - \gamma_5)\,\nu_l\,\bar{c}\,\,\gamma_5\,b \Bigg\} + {\rm h.c.}\,,
\end{eqnarray}
where, 
\begin{eqnarray} 
&&G_V = 1 + {V_L} + {V_R}\,,\qquad\qquad
G_A = 1 + {V_L} - {V_R}\,, \qquad\qquad 
G_S = {S_L} + {S_R}\,,\qquad\qquad
G_P = {S_L} - {S_R}\,. 
\end{eqnarray}
Within the SM, ${V_{L,R}}={S_{L,R}}=0$.

Using the effective Lagrangian of Eq.~(\ref{effl}), the three body differential decay distribution for the $B_1 \to B_2\,l\,\nu$ decays can be written as
\begin{equation}
\frac{d^2\Gamma}{dq^2 d\cos\theta}= \frac{G_{F}^2 |V_{cb}|^2 |\vec{P}_{B_2}|}{2^9\,\pi^3\,m_{B_1}^2} \left(1-\frac{m_{l}^2}{q^2}\right)\,
L_{\mu \nu} H^{\mu \nu}\,,
\end{equation}
where $L_{\mu \nu}$ and $H^{\mu \nu}$ are the leptonic and hadronic current tensors. Here 
$|\vec{P}_{B_2}|= \sqrt{\lambda (m_{B_1}^2, m_{B_2}^2, q^2)}/2m_{B_1}$ with $\lambda(a,b,c)=a^2+b^2+c^2-2(ab+bc+ca)$ represent the three 
momentum vector of the outgoing baryon. One can use the helicity techniques for the 
covariant contraction of $L_{\mu \nu}$ and $H^{\mu \nu}$ details of which can be found in Refs.~\cite{Korner:1989qb,Kadeer:2005aq}. 
We follow Ref.~\cite{Dutta:2013qaa} and write the expression for differential decay distribution for $B_1 \to B_2\,l\,\nu$ decays in 
terms of the helicity amplitudes $H^{V/A}_{\lambda_2\lambda_W}$ are expressed in $\mathcal{A}_1$, $\mathcal{A}_2$, 
$\mathcal{A}_3$, $\mathcal{A}_4$ as follows:
\begin{equation}
\label{dslnutheta}
\frac{d^2\Gamma}{dq^2\,d\cos\theta} = N\left(1-\frac{m^2_l}{q^2}\right)^2\left[\mathcal{A}_1+\frac{m^2_l}{q^2}\mathcal{A}_2+2\mathcal{A}_3+\frac{4m_l}{\sqrt{q^2}}\mathcal{A}_4\right]
\end{equation}
where $\theta$ is the angle between the $\vec{P}_{B_2}$ and lepton three momentum vector in the $l-\nu$ rest frame and
\begin{eqnarray}
N &=& \frac{G_F^2\,|V_{cb}|^2\,q^2|\vec{P}_{B_2}|}{512\,\pi^3\,m_{B_1}^2}\,\Big(1 - \frac{m_l^2}{q^2}\Big)^2,\nonumber\\
\mathcal{A}_1 &=& 2\sin^2\theta\left(H^2_{\frac{1}{2}0}+H^2_{-\frac{1}{2}0}\right)+\left(1-\cos\theta\right)^2H^2_{\frac{1}{2}1}+\left(1+\cos\theta\right)^2H^2_{-\frac{1}{2}-1},\nonumber\\
\mathcal{A}_2 &=& 2\cos^2\theta\left(H^2_{\frac{1}{2}0}+H^2_{-\frac{1}{2}0}\right)+\sin^2\theta\left(H^2_{\frac{1}{2}1}+H^2_{-\frac{1}{2}-1}\right)+2\left(H^2_{\frac{1}{2}t}+H^2_{-\frac{1}{2}t}\right)-4\cos\theta\left(H_{\frac{1}{2}t}H_{\frac{1}{2}0}+H_{-\frac{1}{2}t}H_{-\frac{1}{2}0}\right),\nonumber\\
\mathcal{A}_3 &=& \left(H^{SP}_{\frac{1}{2}0}\right)^2 +\left(H^{SP}_{-\frac{1}{2}0}\right)^2 ,\nonumber\\
\mathcal{A}_4 &=&-\cos\theta\left(H_{\frac{1}{2}0}H^{SP}_{\frac{1}{2}0}+H_{-\frac{1}{2}0}H^{SP}_{-\frac{1}{2}0}\right)+ \left(H_{\frac{1}{2}t}H^{SP}_{\frac{1}{2}0}+H_{-\frac{1}{2}t}H^{SP}_{-\frac{1}{2}0}\right).
\end{eqnarray}

\subsection{Form factors and Helicity amplitudes}
\begin{figure}[htbp]
\centering
\includegraphics[width=8.3cm,height=5.3cm]{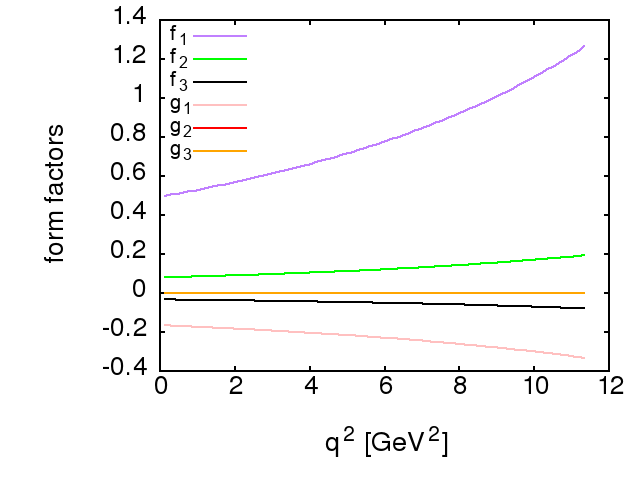}
\includegraphics[width=8.3cm,height=5.3cm]{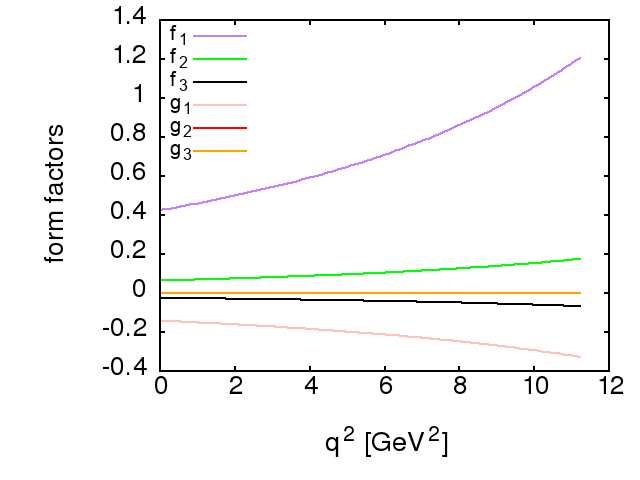}
\caption{$\Sigma_b \to \Sigma_c$~(left) and $\Omega_b \to \Omega_c$~(right) transition form factors as a function of $q^2$.}
\label{fig_ffactor}
\end{figure}

The hadronic matrix elements of vector and axial vector currents between two spin half baryons are parametrized in terms of the following form 
factors
\begin{eqnarray}
M^V_{\mu} &=& \langle B_2,\lambda_2\vert \bar{c}\gamma_{\mu}b\vert B_1,\lambda_1\rangle = \bar{u}_2(p_2,\lambda_2)\left[f_1(q^2)\gamma_{\mu}+if_2(q^2)\sigma_{\mu\nu}q^{\nu}+f_3(q^2)q_{\mu}\right]u_1(p_1,\lambda_1),\nonumber\\
M^A_{\mu} &=& \langle B_2,\lambda_2\vert \bar{c}\gamma_{\mu}\gamma_5 b\vert B_1,\lambda_1\rangle = \bar{u}_2(p_2,\lambda_2)\left[g_1(q^2)\gamma_{\mu}+ig_2(q^2)\sigma_{\mu\nu}q^{\nu}+g_3(q^2)q_{\mu}\right]\gamma_5 u_1(p_1,\lambda_1),
\end{eqnarray}
where $q^{\mu}=(p_1-p_2)^{\mu}$ is the four momentum transfer, $\lambda_1$ and $\lambda_2$ are the respective helicities of the parent and daughter 
baryons and $\sigma_{\mu\nu}=\frac{i}{2}\left[\gamma_{\mu}\gamma_{\nu}\right]$. Here, $B_1$ represents the bottomed baryon $\Sigma_b$ or $\Omega_b$ 
and $B_2$ represents the charmed baryon $\Sigma_c$ or $\Omega_c$. In the heavy quark limit, these matrix elements can be parametrized in terms of 
four velocities $v^{\mu}$ and $v'^{\mu}$ as follows
\begin{eqnarray}
M^V_{\mu} &=& \langle B_2,\lambda_2\vert \bar{c}\gamma_{\mu}b\vert B_1,\lambda_1\rangle = \bar{u}_2(p_2,\lambda_2)\left[F_1(w)\gamma_{\mu}+F_2(w)v_{\mu}+F_3(w)v'_{\mu}\right]u_1(p_1,\lambda_1),\nonumber\\
M^A_{\mu} &=& \langle B_2,\lambda_2\vert \bar{c}\gamma_{\mu}\gamma_5 b\vert B_1,\lambda_1\rangle = \bar{u}_2(p_2,\lambda_2)\left[G_1(w)\gamma_{\mu}+G_2(w)v_{\mu}+G_3(w)v'^{\mu}\right]\gamma_5 u_1(p_1,\lambda_1),
\end{eqnarray}
where $w=v.v'=\left(m^2_{B_1}+m^2_{B_2}-q^2\right)/2m_{B_1}m_{B_2}$ and $m_{B_1}$ and $m_{B_2}$ are the masses of the $B_1$ and $B_2$ baryons, 
respectively. One can compute the hadronic form factors for scalar and pseudo-scalar currents by using the equation of motion. Those matrix elements 
are 
\begin{eqnarray}
\langle B_2,\lambda_2\vert \bar{c}b\vert B_1,\lambda_1\rangle &=& \bar{u}_2(p_2,\lambda_2)\left[f_1(q^2)\frac{q}{m_b-m_c}+f_3(q^2)\frac{q^2}{m_b-m_c}\right]u_1(p_1,\lambda_1),\nonumber\\
\langle B_2,\lambda_2\vert \bar{c}\gamma_5 b\vert B_1,\lambda_1\rangle &=& \bar{u}_2(p_2,\lambda_2)\left[-g_1(q^2)\frac{q}{m_b+m_c}-g_3(q^2)\frac{q^2}{m_b+m_c}\right]\gamma_5 u_1(p_1,\lambda_1),
\end{eqnarray}
where $m_b$ and $m_c$ are the respective masses of $b$ and $c$ quarks calculated at the renormalization scale $\mu =m_b$. These two sets of form 
factor are related through the following relations as given below and the $q^2$ behavior of form factors $f's$ and $g's$ are displayed in Fig.~\ref{fig_ffactor}.
\begin{eqnarray}
f_1(q^2)& =& F_1(q^2)+\left(m_{B_1}+m_{B_2}\right)\left[\frac{F_2(q^2)}{2m_{B_1}}+\frac{F_3(q^2)}{2m_{B_2}}\right],\nonumber\\
f_2(q^2) &=& \frac{F_2(q^2)}{2m_{B_1}}+\frac{F_3(q^2)}{2m_{B_2}},\nonumber\\
f_3(q^2) &=& \frac{F_2(q^2)}{2m_{B_1}}-\frac{F_3(q^2)}{2m_{B_2}},\nonumber\\
g_1(q^2)& =& G_1(q^2)-\left(m_{B_1}-m_{B_2}\right)\left[\frac{G_2(q^2)}{2m_{B_1}}+\frac{G_3(q^2)}{2m_{B_2}}\right],\nonumber\\
g_2(q^2) &=& \frac{G_2(q^2)}{2m_{B_1}}+\frac{G_3(q^2)}{2m_{B_2}},\nonumber\\
g_3(q^2) &=& \frac{G_2(q^2)}{2m_{B_1}}-\frac{G_3(q^2)}{2m_{B_2}}.
\end{eqnarray} 

In the heavy quark limit, the form factors can be expressed in terms of the Isgur-Wise function $\zeta_1(w)$ as follows~\cite{Ebert:2006rp}
\begin{eqnarray}
F_1(w)& =& G_1(w) = -\frac{1}{3}\zeta_1(w),\nonumber\\
F_2(w)& =& F_3(w) = \frac{2}{3}\frac{2}{w+1}\zeta_1(w),\nonumber\\
G_2(w)& =& G_3(w) = 0.
\end{eqnarray}
The explicit expression for $\zeta_1(w)$ is found to be,
\begin{equation}
 \zeta_1(w) = \lim_{m_Q \to \infty} \int \frac{d^3 p}{(2\pi)^3} \Psi_{B_2} \left( \mathbf{p} + 2 \mathbf{\epsilon}_d (p) \sqrt{\frac{w-1}{w+1}} 
 \mathbf{e_\Delta} \right) \Psi_{B_1} (\mathbf{p})
\end{equation}
where $\mathbf{e_\Delta}=\Delta/\sqrt{\Delta^2}$, a unit vector in the direction of $\Delta=M_{B_2} \mathbf{v} - M_{B_1} \mathbf{v}$, and 
$B_1$ and $B_2$ are the parent and daughter baryon respectively.
We refer to Ref.~\cite{Ebert:2006rp} for all the omitted details.

The relation between the hadronic matrix elements and the helicity amplitudes are defined as~\cite{Korner:1989qb,Gutsche:2014zna,Gutsche:2015mxa}
\begin{equation}
H^{V/A}_{\lambda_2\lambda_W} = M^{V/A}_{\mu}(\lambda_2)\epsilon^{\dagger\mu}(\lambda_W),
\end{equation}
where  $\lambda_2$ and $\lambda_W$ are the respective helicities of the daughter baryon and the off-shell $W$ boson. 
In the rest frame of parent baryon $B_1$, the helicity amplitudes can be written as~\cite{Shivashankara:2015cta,Dutta:2018zqp}
\begin{eqnarray}
H^V_{\frac{1}{2}0} &=& G_V\frac{\sqrt{Q_{-}}}{\sqrt{q^2}}\left[\left(m_{B_1}+m_{B_2}\right)f_1(q^2)-q^2f_2(q^2)\right],\nonumber\\
H^A_{\frac{1}{2}0} &=& G_A\frac{\sqrt{Q_{+}}}{\sqrt{q^2}}\left[\left(m_{B_1}-m_{B_2}\right)g_1(q^2)+q^2g_2(q^2)\right],\nonumber\\
H^V_{\frac{1}{2}1} &=& G_V\sqrt{2Q_{-}}\left[-f_1(q^2)+\left(m_{B_1}+m_{B_2}\right)f_2(q^2)\right],\nonumber\\
H^A_{\frac{1}{2}1} &=& G_A\sqrt{2Q_{+}}\left[-g_1(q^2)-\left(m_{B_1}-m_{B_2}\right)g_2(q^2)\right],\nonumber\\\
H^V_{\frac{1}{2}t} &=& G_V\frac{\sqrt{Q_{+}}}{\sqrt{q^2}}\left[\left(m_{B_1}-m_{B_2}\right)f_1(q^2)+q^2f_3(q^2)\right],\nonumber\\
H^A_{\frac{1}{2}t} &=& G_A\frac{\sqrt{Q_{-}}}{\sqrt{q^2}}\left[\left(m_{B_1}+m_{B_2}\right)g_1(q^2)-q^2g_3(q^2)\right],
\end{eqnarray}
where $Q_{\pm}=\left(m_{B_1}\pm m_{B_2}\right)^2-q^2$. 
For the helicity flipped components, these amplitudes turn out to be $H^V_{-\lambda_2-\lambda_W}= H^V_{\lambda_2\lambda_W}$ and 
$H^A_{-\lambda_2-\lambda_W}= -H^A_{\lambda_2\lambda_W}$. Hence, the total left-handed helicity amplitude is 
\begin{equation}
H_{\lambda_2\lambda_W} = H^V_{\lambda_2\lambda_W}-H^A_{\lambda_2\lambda_W}
\end{equation}
The scalar/pseudoscalar helicity amplitudes are defined as
\begin{eqnarray}
H^{SP}_{\frac{1}{2}0} &=& H^{S}_{\frac{1}{2}0}-H^{P}_{\frac{1}{2}0},\nonumber\\
H^{S}_{\frac{1}{2}0} & =& G_S\frac{\sqrt{Q_+}}{m_b-m_c}\left[\left(m_{B_1}-m_{B_2}\right)f_1(q^2)+q^2f_3(q^2)\right],\nonumber\\
H^{P}_{\frac{1}{2}0} & =& G_S\frac{\sqrt{Q_-}}{m_b+m_c}\left[\left(m_{B_1}+m_{B_2}\right)g_1(q^2)-q^2g_3(q^2)\right].
\end{eqnarray}
For these amplitudes, the helicity flipped counterparts are $H^S_{-\lambda_2-\lambda_W}= H^S_{\lambda_2\lambda_W}$ and $H^P_{-\lambda_2-\lambda_W}= -H^P_{\lambda_2\lambda_W}$.

\subsection{Decay distribution and $q^2$ observables}
To obtain the normalized differential decay rate, we perform the $\cos\theta$ integration in Eq.~(\ref{dslnutheta}), i.e, 
\begin{equation}
\frac{d\Gamma}{dq^2}=\frac{8N}{3}\left(1-\frac{m^2_l}{q^2}\right)^2\left[\mathcal{B}_1+\frac{m^2_l}{2q^2}\mathcal{B}_2+\frac{3}{2}\mathcal{B}_3+\frac{3m_l}{\sqrt{q^2}}\mathcal{B}_4\right],
\end{equation}
where
\begin{eqnarray}
\mathcal{B}_1 & =& H^2_{\frac{1}{2}0}+H^2_{-\frac{1}{2}0} +H^2_{\frac{1}{2}1}+H^2_{-\frac{1}{2}-1},\nonumber\\
\mathcal{B}_2 & =& H^2_{\frac{1}{2}0}+H^2_{-\frac{1}{2}0} +H^2_{\frac{1}{2}1}+H^2_{-\frac{1}{2}-1}+3\left(H^2_{\frac{1}{2}t}+H^2_{-\frac{1}{2}t}\right),\nonumber\\
\mathcal{B}_3 & =& \left(H^{SP}_{\frac{1}{2}0}\right)^2 +\left(H^{SP}_{-\frac{1}{2}0}\right)^2 ,\nonumber\\
\mathcal{B}_4 &=& H_{\frac{1}{2}t}H^{SP}_{\frac{1}{2}0}+H_{-\frac{1}{2}t}H^{SP}_{-\frac{1}{2}0}.
\end{eqnarray}

The SM equations can be obtained by setting $G_V=G_A=1$ and $\widetilde{G}_V = \widetilde{G}_A = 0$. 

The ratio of branching ratio which is defined by considering the ratios of the differential decay rate having the heavier $\tau$ lepton in the final
state to the differential decay rate having the lighter lepton in the final state as,
\begin{eqnarray}
R_{B_2} = \frac{\Gamma(B_1 \to B_2\,\tau\,\nu)}{\Gamma(B_1 \to B_2\,l\,\nu)}\,,
\end{eqnarray}
where $B_{1(2)} = \Sigma_{b(c)}, \Omega_{b(c)}$ and $l=e$ or $\mu$.

Similarly, we also define various $q^2$ dependent observables such as total differential decay rate ${d\Gamma}/{dq^2}(q^2)$, 
ratio of branching ratio $R_{B_2}(q^2)$, forward backward asymmetry $A_{FB}^l(q^2)$ obtained by integrating over linear $cos\, \theta$ dependency
of the distribution, 
polarization fraction of the charged lepton $P^{l}(q^2)$ calculated by measuring the difference between the lepton helicity nonflip rate
to the lepton helicity flip rate and convexity parameter $C_{F}^l (q^2)$ which is found by integrating over $cos^2\, \theta$ dependency
of the distribution for both the decay modes as follows:
\begin{eqnarray}
&&R_{B_2}(q^2)=\frac{\Gamma(B_1 \rightarrow B_2 \tau \nu)}{\Gamma(B_1 \rightarrow B_2\,l\,\nu)}\,, \qquad\qquad
A_{FB}^{l}(q^2) = \frac{\Big(\int_{-1}^{0}-\int_{0}^{1}\Big)d\cos\theta\frac{d^2\Gamma}{dq^2\,d\cos\theta}}{\frac{d\Gamma}{dq^2}} \,, \nonumber \\
&&P^l(q^2)=\frac{d\Gamma(+)/dq^2 - d\Gamma(-)/dq^2}{d\Gamma(+)/dq^2 + d\Gamma(-)/dq^2}\,, \qquad
C_{F}^l(q^2)= \frac{1}{\left(d\Gamma/dq^2\right)} \frac{d^2}{d(\cos\theta)^2}\left[\frac{d^2\Gamma}{dq^2\,d\cos\theta}\right]\,,
\end{eqnarray}
where $d\Gamma(+)/dq^2$ and $d\Gamma(-)/dq^2$ are the respective differential decay rates of positive and negative helicity of lepton.

\section{Results and Discussions} 
\label{results}
\subsection{Input Parameter}
For our numerical computation of various observables we use the input parameters from Ref.~\cite{Tanabashi:2018oca} and, for definiteness,
we report it in Table~\ref{tab_inputs}. Masses of all the particles are in ${\rm GeV}$ units and Fermi coupling constant $G_F$ is in
${\rm GeV^{-2}}$ unit. For the 
$\Sigma_b \to \Sigma_c$ and $\Omega_b \to \Omega_c$ transition form factors, we follow Ref.~\cite{Ebert:2006rp} and use the form factor 
inputs obtained in the framework of relativistic quark model. In the 
heavy quark limit the invariant form factors are expressed in terms of the Isgur-Wise functions $\zeta_1(w)$ and $\zeta_2(w)$ 
obtained for the whole kinematic range using the $\Psi_{\Sigma_{(b,c)}}$ and $\Psi_{\Omega_{(b,c)}}$ baryon wave functions. The values
of $\zeta_1(w)$ and $\zeta_2(w)$ in the whole kinematic range, pertinent for our analysis, was obtained from Ref.~\cite{private_comm}.

\begin{table}[htbp]
\centering
\setlength{\tabcolsep}{8pt} 
\renewcommand{\arraystretch}{1.5} 
\begin{ruledtabular}
\begin{tabular}{cccccccc}
    Parameter & Value & Parameter & Value & Parameter & Value & Parameter & Value\\
    \hline
    $m_{\Sigma_b}$ & 5.8155 & $m_{\Sigma_c}$ & 2.45375 & $m_b(m_b)$ & 4.18 & $m_c(m_b)$ & 0.91 \\
    $m_{\Omega_b}$ & 6.0461 & $m_{\Omega_c}$ & 2.6952 & $G_F$ & $1.1663787\times 10^{-5}$ & $|V_{cb}|$ & 0.041(11) \\ 
    $m_e$ & $0.51099\times 10^{-3}$ & $m_{\tau}$ & 1.77682 \\
\end{tabular}
\caption{Theory input parameters~\cite{Tanabashi:2018oca}}
\label{tab_inputs}
\end{ruledtabular}
\end{table}    

\subsection{Standard model predictions}
The SM predictions are reported for $\Sigma_b \to \Sigma_c l \nu$ and $\Omega_b \to \Omega_c l \nu$ decay modes undergoing $b \to cl\nu$ 
quark level transitions where, $l$ is either an electron or a tau lepton.  
In Table~\ref{tab_smcentral}, we display the average values of various observables such as the total decay rate $\Gamma$, longitudinal polarization of the 
charged lepton $\langle P^l \rangle$, forward-backward asymmetry $\langle A_{FB}^l \rangle$, the convexity parameter $\langle C_{F}^l \rangle$ 
for both electron mode and tau mode respectively. We also report the ratio of branching ratios for these decay modes. The total decay rate 
for both the decay modes is 
observed to be larger for the lighter leptons ($e$ or $\mu$) as compared to the heavier $\tau$ lepton. The polarization fraction for
the electron is $-1.00$. The $\tau$ polarization fraction is $0.131$ for $\Sigma_b \to \Sigma_c$ and $0.135$ for $\Omega_b \to \Omega_c$ decay 
modes. The forward-backward asymmetry for electron mode and tau mode are almost similar for both the decay modes. 
The convexity parameter $C_F^l$ for $\tau$ mode is larger than the $e$ mode. The ratio of branching ratio for $\Omega_b \to \Omega_c l \nu$ 
is slightly larger than the $\Sigma_b \to \Sigma_c l \nu$ decay mode. 

We also determine the size of uncertainties in each observable that are coming from various input parameters.
The uncertainties for the theoretical predictions can come from the nonperturbative hadronic form factors and not very well know CKM matrix
element $|V_{cb}|$. Here we consider the form factor uncertainties within $10\%$ and the $|V_{cb}|$ uncertainty as mentioned in Table~\ref{tab_inputs}. 
In order to measure the size of uncertainty, we perform a random scan over the input parameters within $1\sigma$. Graphically the
SM uncertainties for each $q^2$ dependant observable are displayed in Fig.~\ref{fignpsig} and \ref{fignpomg} with a red patch.
Interestingly, the SM uncertainties are very small in all the observables except for the total differential decay rate $d\Gamma/dq^2$.

\begin{table}[htbp]
\centering
\setlength{\tabcolsep}{8pt} 
\renewcommand{\arraystretch}{1.5} 
\begin{tabular}{|c|c|c|c|c|}
\hline
\hline
     &\multicolumn{2}{c|}{$\Sigma_b \to \Sigma_c l \nu$ } &\multicolumn{2}{c|}{$\Omega_b \to \Omega_c l \nu$} \\
     \cline{2-5}
     &$e$ mode&$\tau$ mode&$e$ mode&$\tau$ mode\\
    \hline
    \hline
     $ \Gamma \times 10^{10}$ s$^{-1}$& 1.401 & 0.473 & 1.235 & 0.447\\
    \hline
     $\langle P^l \rangle$& -1.000 & 0.131 & -1.000 & 0.135  \\
    \hline
     $\langle A_{FB}^l \rangle$& 0.050 & -0.253 & 0.050 & -0.251  \\
    \hline
     $\langle C_{F}^{l} \rangle$& -1.172 & -0.200 & -1.148 & -0.196  \\
    \hline
    $\langle R \rangle$&\multicolumn{2}{c|}{$R_{\Sigma_c}=0.338$} &\multicolumn{2}{c|}{$R_{\Omega_c}=0.362$} \\
\hline
\hline
\end{tabular}
\caption{The SM central values for the ratio of branching ratio $\langle R \rangle$, 
the total decay rate $\Gamma$, the lepton polarization fraction 
$\langle P^l \rangle$, the forward-backward asymmetry $\langle A_{FB}^l \rangle$ and the convexity factor $\langle C_{F}^{l} \rangle$
for the $e$ mode and the $\tau$ mode of $\Sigma_b \to \Sigma_c l \nu$ and $\Omega_b \to \Omega_c l \nu$ decays.}
\label{tab_smcentral}
\end{table}

\begin{figure}[ht]
\centering
\includegraphics[width=5.4cm,height=4.0cm]{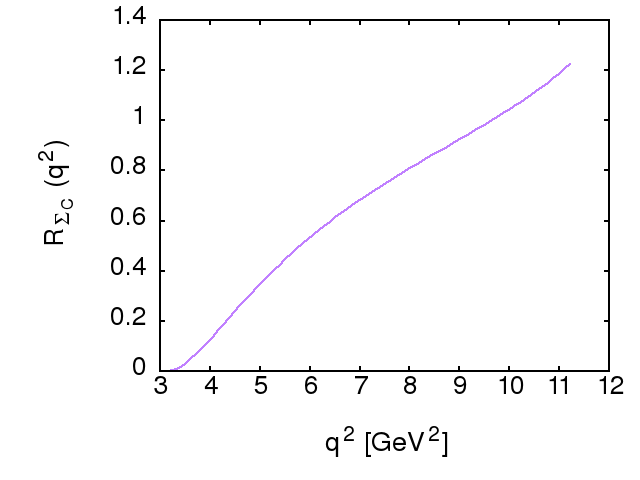}
\includegraphics[width=5.4cm,height=4.0cm]{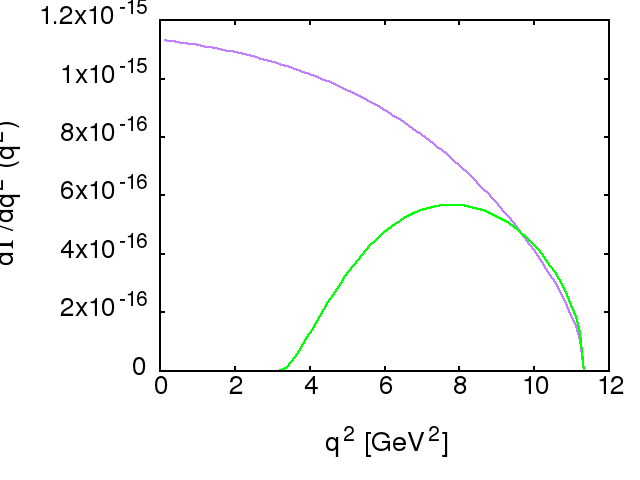}
\includegraphics[width=5.4cm,height=4.0cm]{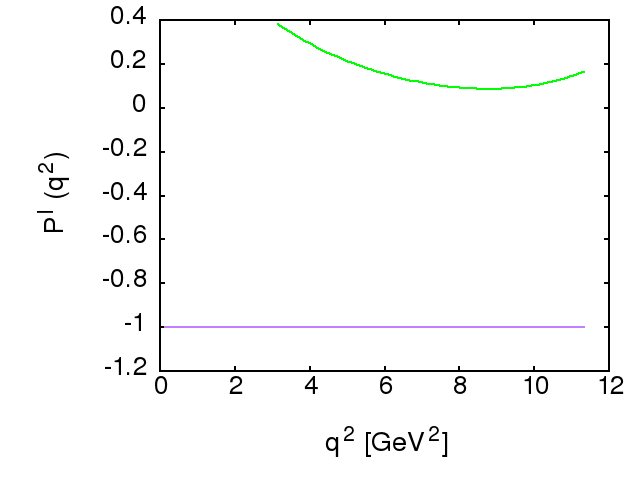}
\includegraphics[width=5.4cm,height=4.0cm]{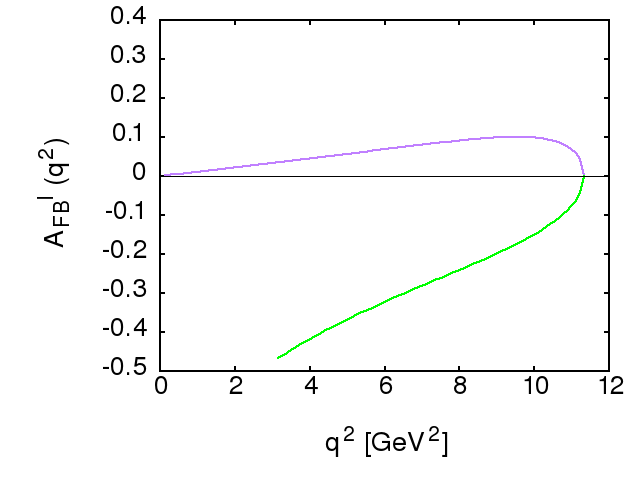}
\includegraphics[width=5.4cm,height=4.0cm]{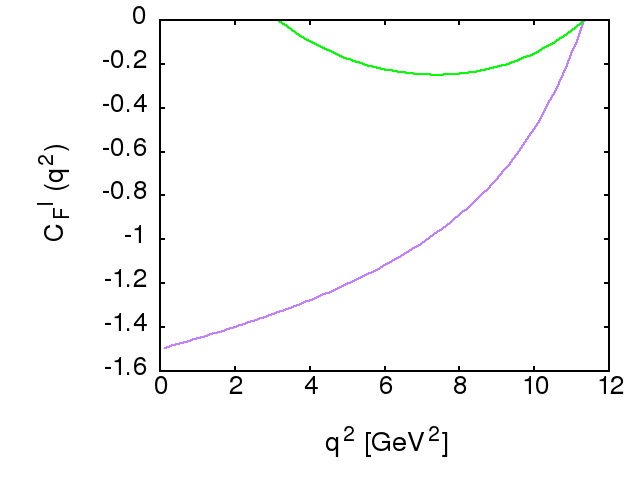}
\caption{Ratio of branching ratio $R_{\Sigma_c}(q^2)$, 
the total differential decay rate $d\Gamma/dq^2$, the lepton polarization fraction 
$P^{l}(q^2)$, the forward-backward asymmetry $A_{FB}^{l}(q^2)$ and the convexity parameter $C_{F}^{l}(q^2)$ for the 
$\Sigma_b \to \Sigma_c l \nu$ decays in the SM.
The purple color represents the $e$ mode and the green color represents the $\tau$ mode. }
\label{fig_smsig}
\end{figure}

\begin{figure}[htbp]
\centering
\includegraphics[width=5.4cm,height=4.0cm]{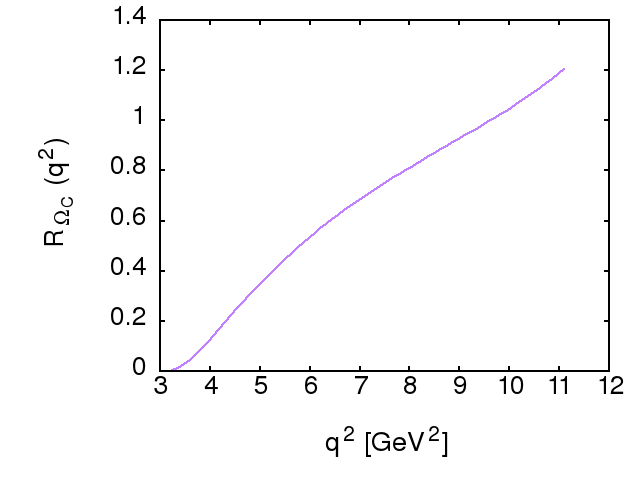}
\includegraphics[width=5.4cm,height=4.0cm]{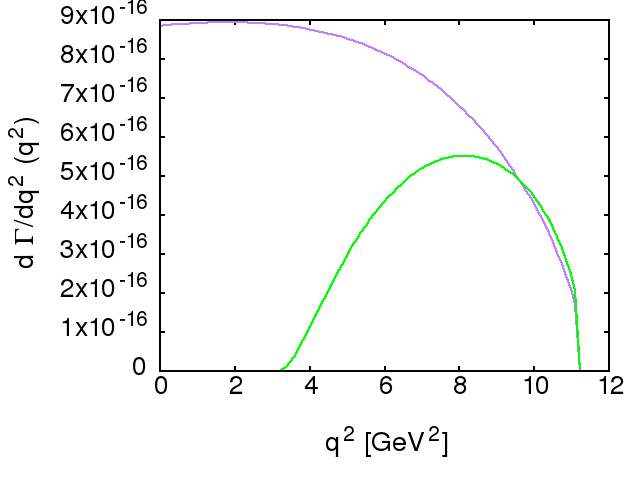}
\includegraphics[width=5.4cm,height=4.0cm]{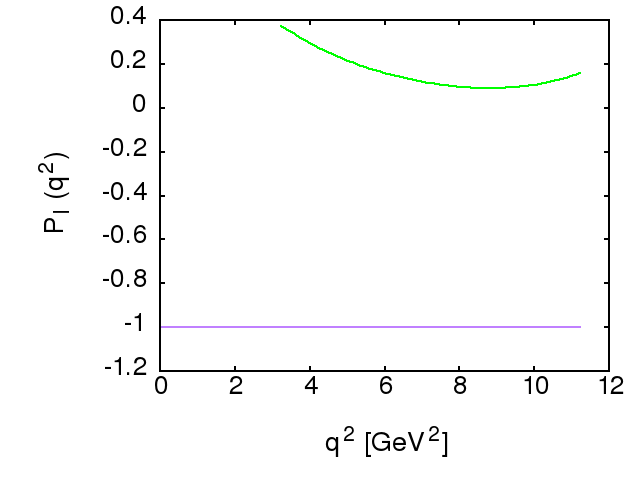}
\includegraphics[width=5.4cm,height=4.0cm]{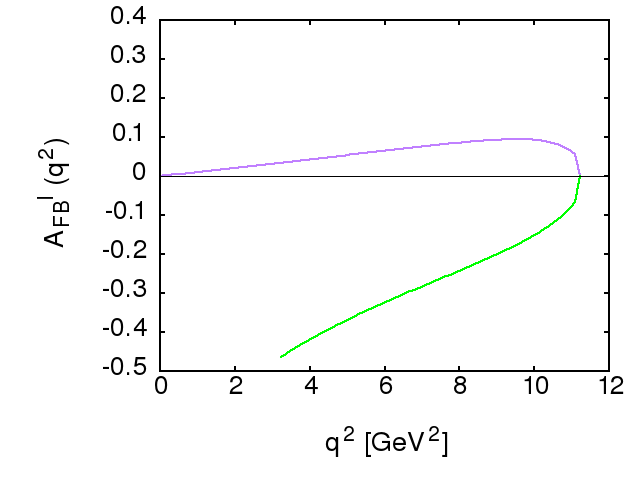}
\includegraphics[width=5.4cm,height=4.0cm]{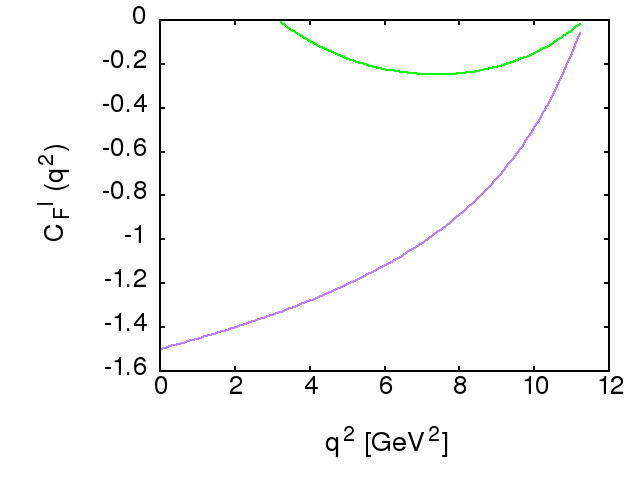}
\caption{Ratio of branching ratio $R_{\Omega_c}(q^2)$, 
the total differential decay rate $d\Gamma/dq^2$, the lepton polarization fraction 
$P^{l}(q^2)$, the forward-backward asymmetry $A_{FB}^{l}(q^2)$ and the convexity parameter $C_{F}^{l}(q^2)$ as a function of $q^2$
for the $\Omega_b \to \Omega_c l \nu$ decays in the SM.
The purple color represents the $e$ mode and the green color represents the $\tau$ mode.}
\label{fig_smomg}
\end{figure}

The behavior of each observable as a function of $q^2$ for $\Sigma_b \to \Sigma_c l \nu$ and $\Omega_b \to \Omega_c l \nu$ decays
are reported in the Figure~\ref{fig_smsig} and \ref{fig_smomg}. 
We compare each observable for both electron and tau lepton final states.
The purple color represents the electron mode and the green color represents the tau
mode. The $q^2$ dependence of all the observable are distinct for both $e$ and $\tau$ modes. 
The $R_{\Sigma_c}(q^2)$ show almost positive slope over the entire $q^2$ range. The total differential decay rate for electron is maximum
at minimum $q^2$ and minimum at maximum $q^2$ whereas, the the total differential decay rate for tau is maximum
at around $q^2=8$ GeV$^2$ and approaches zero at minimum and maximum $q^2$. 
The $P^e(q^2)$ is -1 over entire $q^2$ range and the $P^{\tau}(q^2)$ take only positive values for all $q^2$ values.
The $A_{FB}^l(q^2)$ is positive in $e$ mode while it is negative in $\tau$ mode in the whole $q^2$ range. At $q^2=q^2_{max}$, both 
$A_{FB}^e(q^2)$ and $A_{FB}^{\tau}(q^2)$ approaches to zero. The $C_{F}^{e}(q^2)$ is around -1.5 at $q^2=m_l^2$ and zero at maximum $q^2$.
On the other hand $C_{F}^{\tau}(q^2)$ approaches zero at both minimum and maximum $q^2$.
Similar conclusions can be made for $\Omega_b \to \Omega_c l \nu$ decay mode as well.

\subsection{New physics analysis}
\begin{figure}[htbp]
\centering
\includegraphics[width=5.4cm,height=4.0cm]{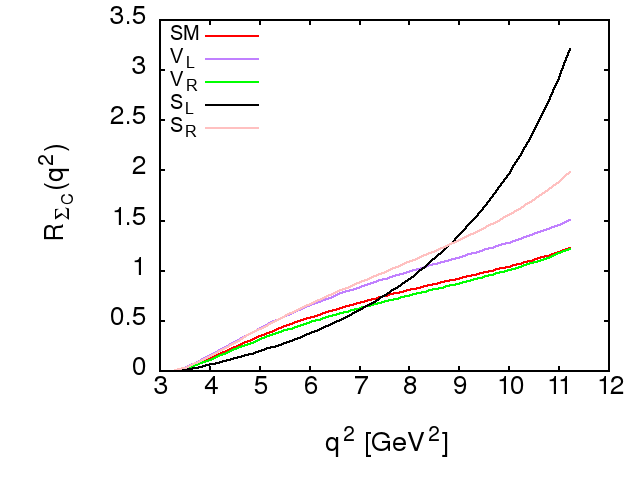}
\includegraphics[width=5.4cm,height=4.0cm]{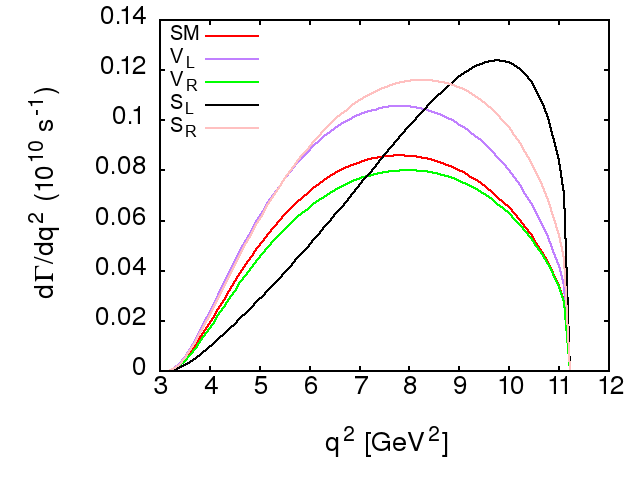}
\includegraphics[width=5.4cm,height=4.0cm]{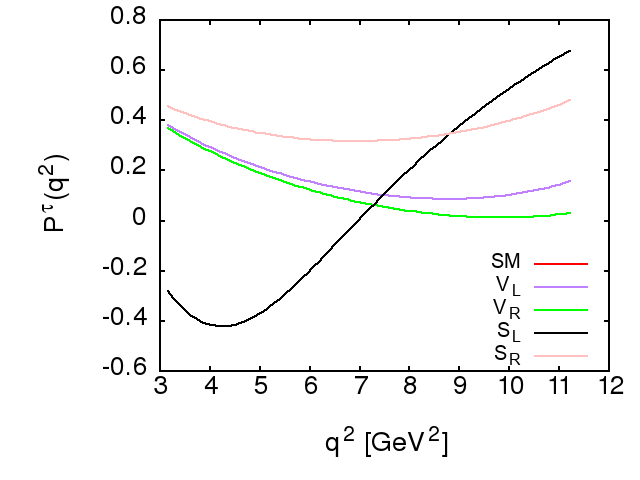}
\includegraphics[width=5.4cm,height=4.0cm]{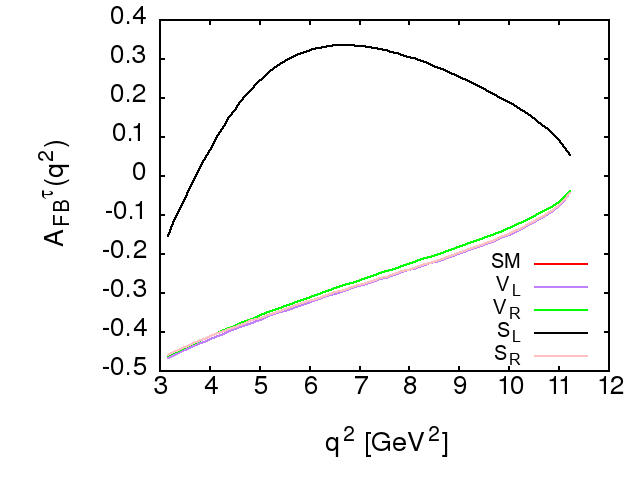}
\includegraphics[width=5.4cm,height=4.0cm]{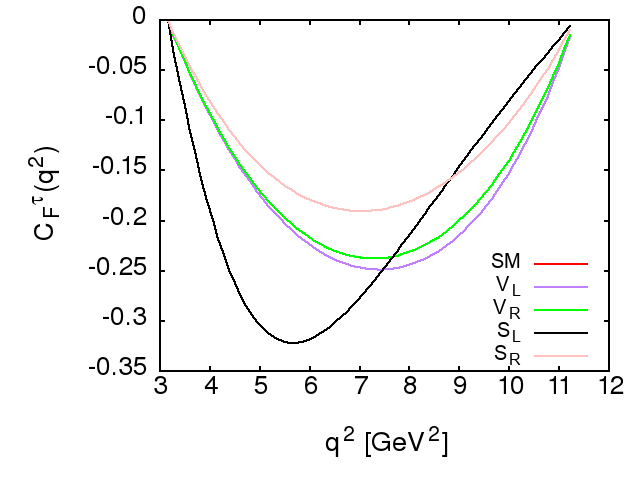}
\caption{The $q^2$ dependency of the ratio of branching ratio $R_{\Sigma_c}(q^2)$, 
the total differential decay rate $d\Gamma/dq^2$, the lepton polarization fraction 
$P^{\tau}(q^2)$, the forward-backward asymmetry $A_{FB}^{\tau}(q^2)$ and the convexity factor $C_{F}^{\tau}(q^2)$ in SM (red) and in the presence of
$V_L$ (purple), $V_R$ (green), $S_L$ (black), $S_R$ (pink) NP couplings for the $\Sigma_b \to \Sigma_c \tau \nu$ decay mode.}
\label{fig_everysig}
\end{figure}

\begin{figure}[htbp]
\centering
\includegraphics[width=5.4cm,height=4.0cm]{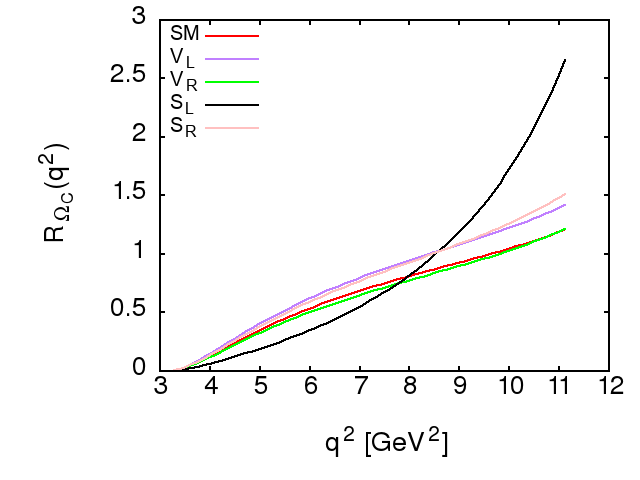}
\includegraphics[width=5.4cm,height=4.0cm]{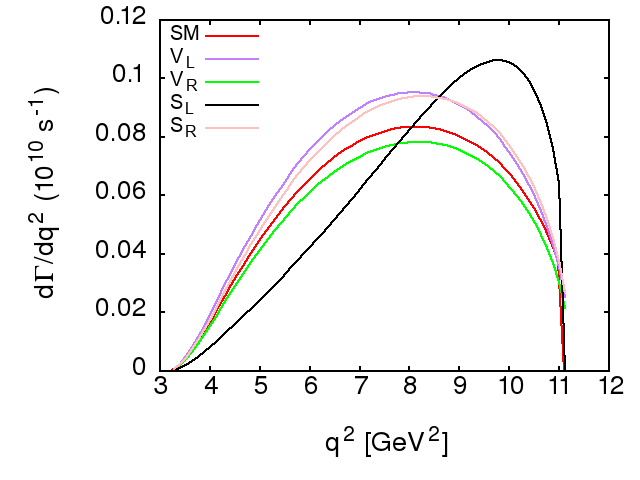}
\includegraphics[width=5.4cm,height=4.0cm]{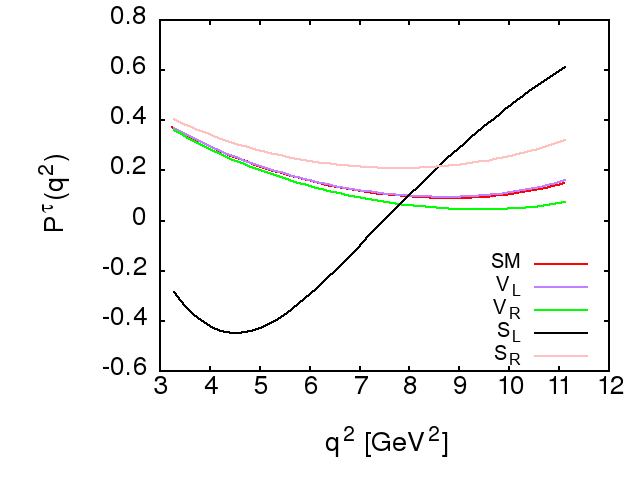}
\includegraphics[width=5.4cm,height=4.0cm]{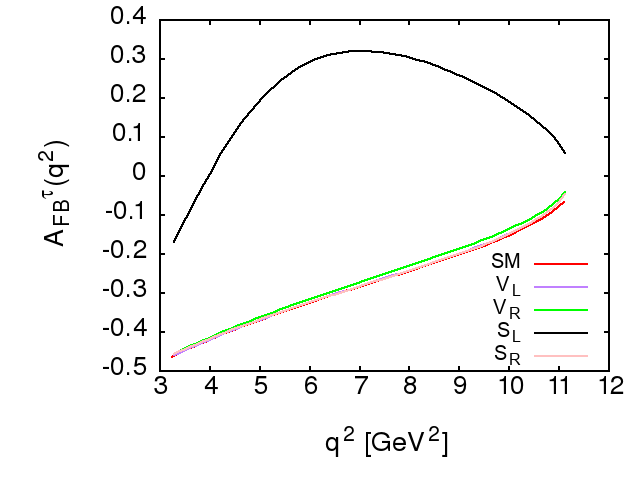}
\includegraphics[width=5.4cm,height=4.0cm]{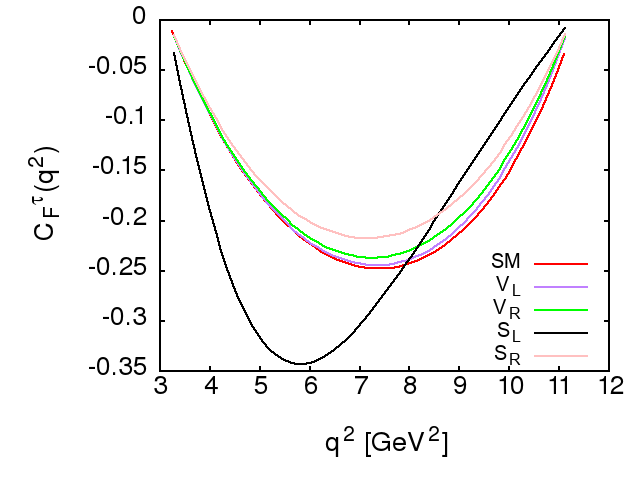}
\caption{The $q^2$ dependency of the ratio of branching ratio $R_{\Omega_c}(q^2)$, 
the total differential decay rate $d\Gamma/dq^2$, the lepton polarization fraction 
$P^{\tau}(q^2)$, the forward-backward asymmetry $A_{FB}^{\tau}(q^2)$ and the convexity factor $C_{F}^{\tau}(q^2)$ in SM (red) and in the presence of
$V_L$ (purple), $V_R$ (green), $S_L$ (black), $S_R$ (pink) NP couplings for the $\Omega_b \to \Omega_c \tau \nu$ decay mode.}
\label{fig_everyomg}
\end{figure}

We anlayse the NP effects in a model independent way.
The new physics effects are investigated in four different scenarios by considering each new vector and scalar type NP couplings 
associated with the left handed neutrinos one at a time. The effects of $V_L$, $V_R$, $S_L$ and $S_R$ NP couplings are studied for both  
$\Sigma_b \to \Sigma_c \tau \nu$ and $\Omega_b \to \Omega_c \tau \nu$ decay modes. To get the allowed NP parameter space in each scenario, 
we impose the $3\sigma$ constraint coming from the measured values of the ratio of branching ratios $R_D$ and $R_{D^*}$. 
We first perform a naive $\chi^2$ test to find the best fit values of each observable by defining
\begin{equation}
\label{totchisq}
\left[\chi^{2}\right]_{Total} = \frac{\left[R_{D}^{expt} - R_{D}^{th}\right]^2}{\left[\Delta R_{D}^{expt} \right]^2} + 
                                \frac{\left[R_{D^*}^{expt} - R_{D^*}^{th}\right]^2}{\left[\Delta R_{D^*}^{expt} \right]^2}
\end{equation}
where, $R_{D}^{expt}$ and $R_{D^*}^{expt}$ refer to the experimental values of $R_D$ and $R_{D^*}$ and 
$\Delta R_{D}^{expt}$, $\Delta R_{D^*}^{expt}$ refer to the experimental uncertainties associated with $R_D$ and $R_{D^*}$ and 
similarly $R_{D}^{th}$, $R_{D^*}^{th}$ refer to the theoretical values corresponding to various NP couplings. For the uncertainties in
$R_D$ and $R_{D^{\ast}}$, we added the systematic and statistical uncertainties in quadrature.
To calculate the best fit values, we evaluate the minimum $\chi^2$ and find the respective best fit values for each $V_L$, $V_R$, $S_L$ and 
$S_R$ NP couplings. In Table~\ref{tab_bestfit}, we display the corresponding best fit average values of each observable 
associated with $V_L$, $V_R$, $S_L$ and $S_R$ NP couplings for the $\Sigma_b \to \Sigma_c \tau \nu$ and $\Omega_b \to \Omega_c \tau \nu$ 
decay modes. 
Although, there are deviations of each observable in each NP scenarios, the forward backward asymmetry $\langle A_{FB}^{\tau}\rangle$ 
corresponding to $S_L$ shows completely different pattern for both
$\Sigma_b \to \Sigma_c \tau \nu$ and $\Omega_b \to \Omega_c \tau \nu$ decay modes. It assumes positive values for $S_L$ and negative for the
rest of the NP couplings. Measurement of $\langle A_{FB}^{\tau}\rangle$ for these decay modes in future will be crucial in distinguishing
various NP Lorentz structures.
We also compare in Fig.~\ref{fig_everysig} and Fig.~\ref{fig_everyomg} various $q^2$ dependent observables obtained using the 
best fit values of each NP couplings with the SM central value. It is evident that the deviation observed with $S_L$ NP coupling is quite 
different from all the other NP couplings in both the decay modes. 
\begin{table}[htbp]
\centering
\setlength{\tabcolsep}{8pt} 
\renewcommand{\arraystretch}{1.5} 
\begin{tabular}{|c|c|c|c|c|c|c|c|c|}
\hline
\hline
     &\multicolumn{4}{c|}{$\Sigma_b \to \Sigma_c \tau \nu$ } &\multicolumn{4}{c|}{$\Omega_b \to \Omega_c \tau \nu$} \\
     \cline{2-9}
     
     &$V_L$ &$V_R$ &$S_L$ &$S_R$ &$V_L$ &$V_R$ &$S_L$ &$S_R$\\
    \hline
    \hline
     $ \Gamma \times 10^{10}$ s$^{-1}$& 0.548 & 0.450 & 0.489 & 0.538 & 0.518 & 0.426 & 0.466 & 0.509 \\
    \hline
     $\langle P^{\tau}\rangle$        & 0.131 & 0.092 & 0.159 & 0.236 & 0.135 & 0.095 & 0.170 & 0.241 \\
    \hline
     $\langle A_{FB}^{\tau}\rangle$   & -0.253 & -0.241 & 0.242 & -0.250 & -0.251 & -0.239 & 0.240 & -0.248 \\
    \hline
     $\langle C_{F}^{\tau}\rangle$    & -0.200 & -0.192 & -0.193 & -0.176 & -0.196 & -0.189 & -0.188 & -0.172 \\
    \hline
    $\langle R \rangle$               & 0.391 & 0.321 & 0.349 & 0.384 & 0.419 & 0.345 & 0.377 & 0.421 \\
\hline
\hline  
\end{tabular}
\caption{Ratio of branching ratio $\langle R \rangle$, the total decay rate $\Gamma$, the tau polarization fraction 
$\langle P^{\tau} \rangle$, the forward-backward asymmetry $\langle A_{FB}^{\tau} \rangle$ and the convexity parameter
$\langle C_{F}^{\tau} \rangle$
for $\Sigma_b \to \Sigma_c \tau \nu$ and $\Omega_b \to \Omega_c \tau \nu$ decay modes with the best fit value of each NP couplings.}
\label{tab_bestfit}
\end{table}

\begin{figure}[htbp]
\centering
\includegraphics[width=4.3cm,height=3.3cm]{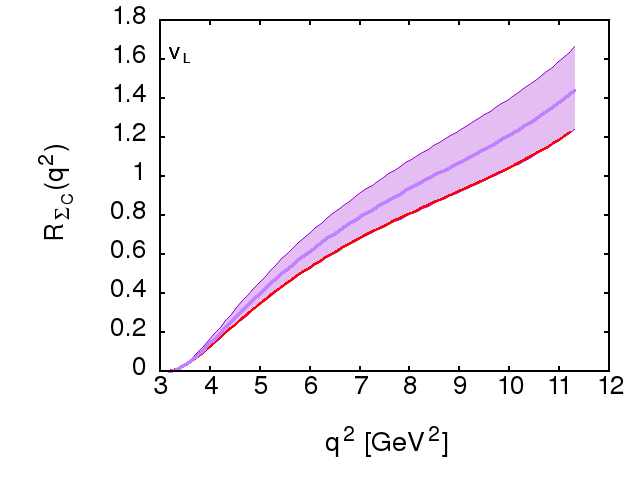}
\includegraphics[width=4.3cm,height=3.3cm]{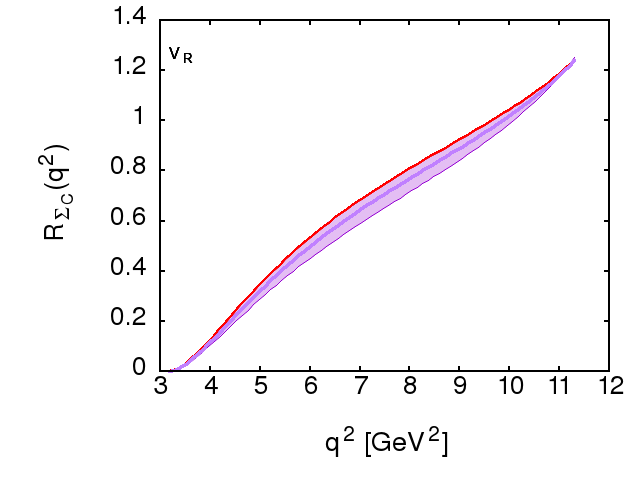}
\includegraphics[width=4.3cm,height=3.3cm]{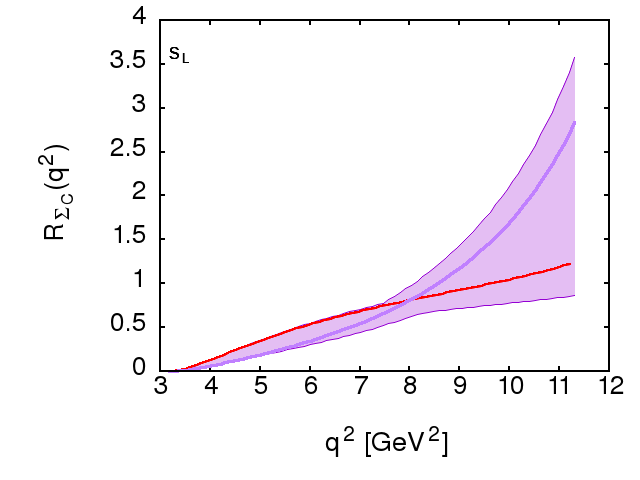}
\includegraphics[width=4.3cm,height=3.3cm]{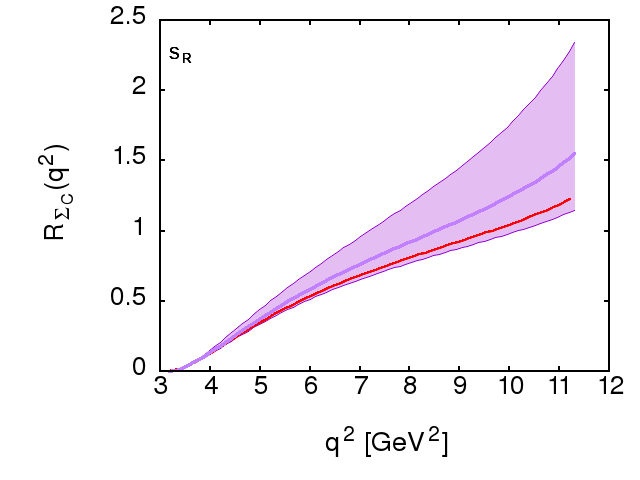}
\includegraphics[width=4.3cm,height=3.3cm]{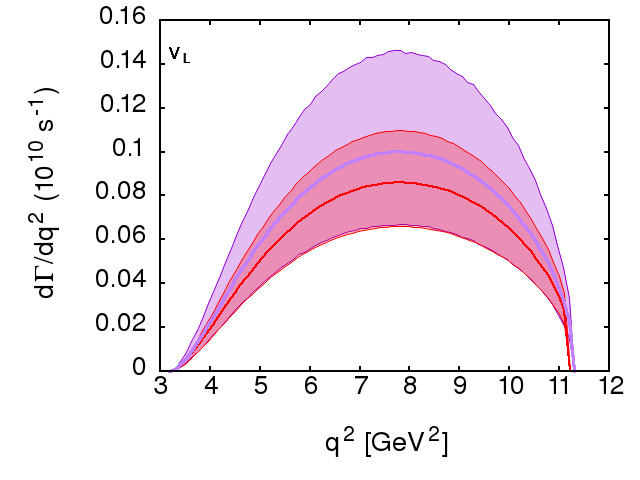}
\includegraphics[width=4.3cm,height=3.3cm]{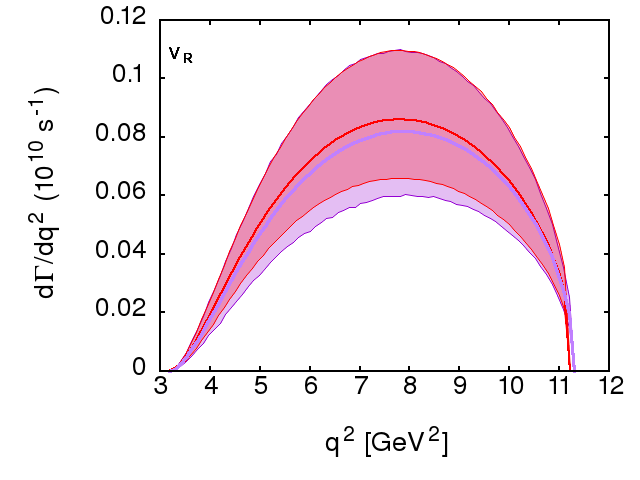}
\includegraphics[width=4.3cm,height=3.3cm]{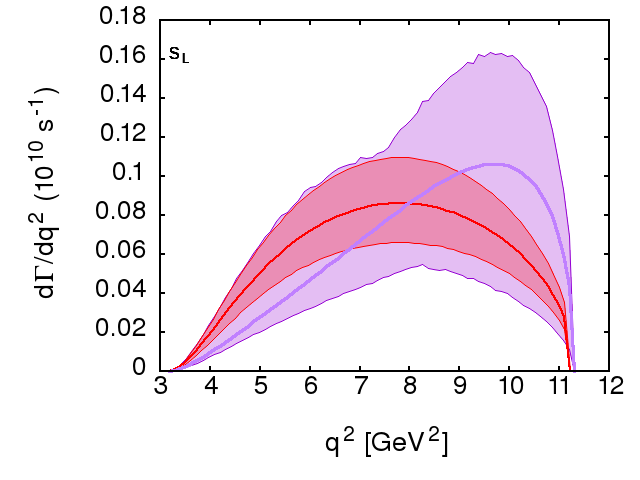}
\includegraphics[width=4.3cm,height=3.3cm]{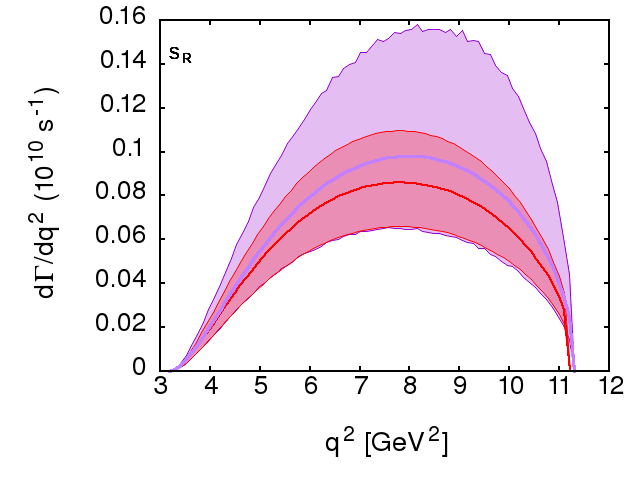}
\includegraphics[width=4.3cm,height=3.3cm]{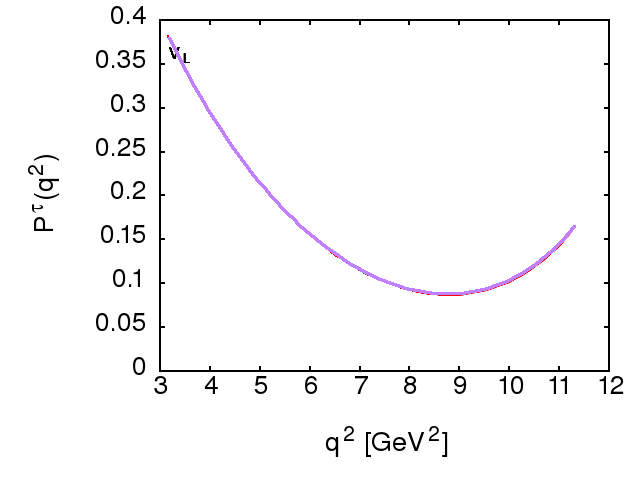}
\includegraphics[width=4.3cm,height=3.3cm]{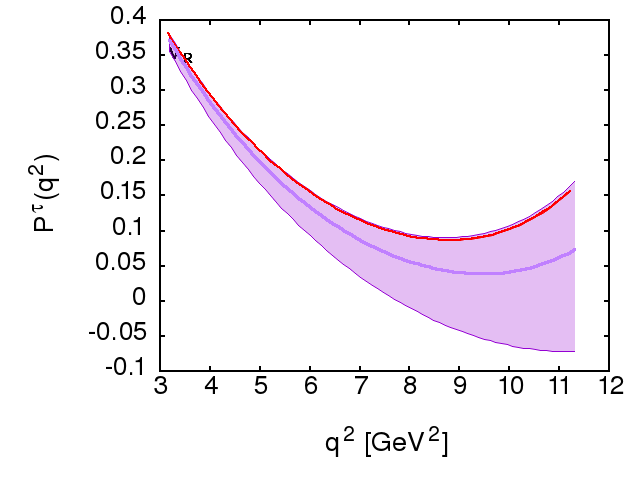}
\includegraphics[width=4.3cm,height=3.3cm]{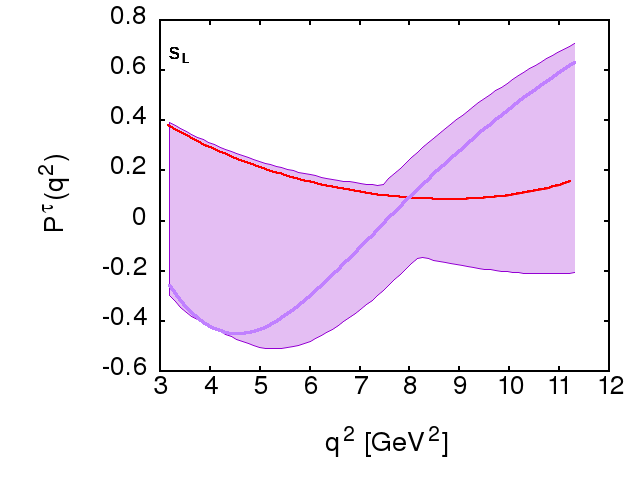}
\includegraphics[width=4.3cm,height=3.3cm]{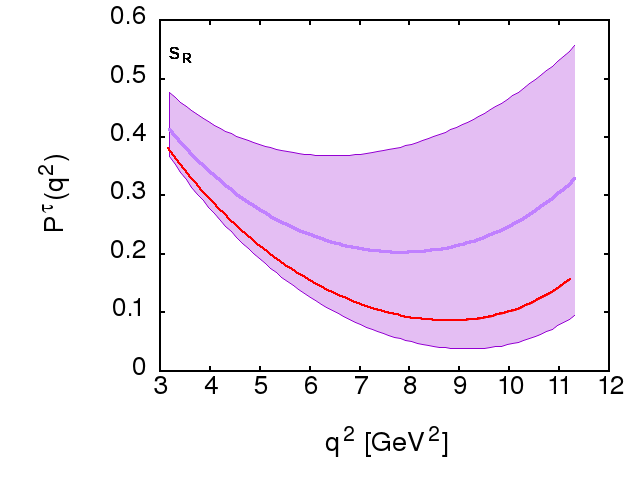}
\includegraphics[width=4.3cm,height=3.3cm]{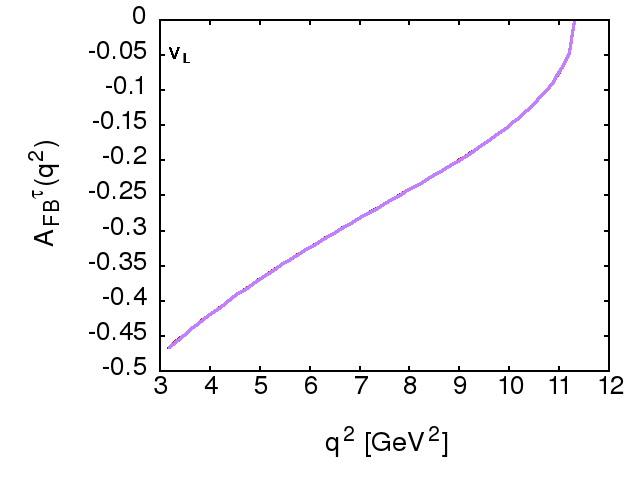}
\includegraphics[width=4.3cm,height=3.3cm]{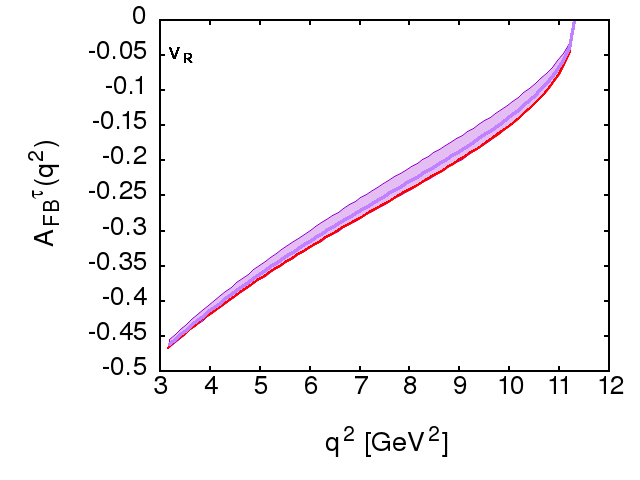}
\includegraphics[width=4.3cm,height=3.3cm]{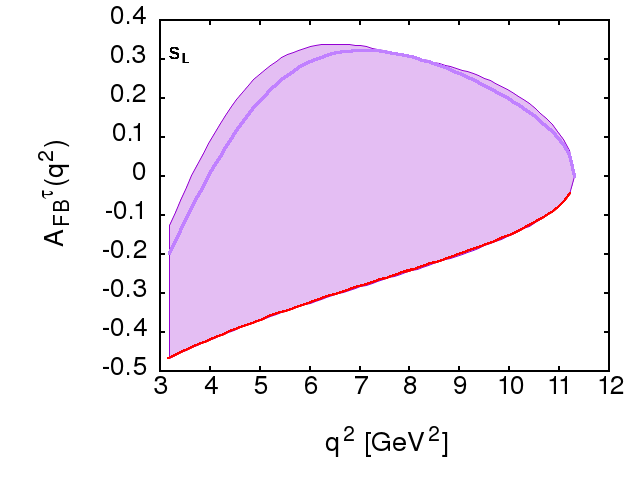}
\includegraphics[width=4.3cm,height=3.3cm]{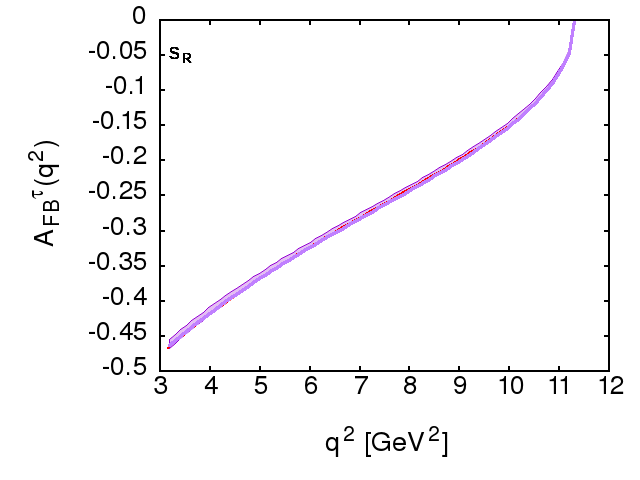}
\includegraphics[width=4.3cm,height=3.3cm]{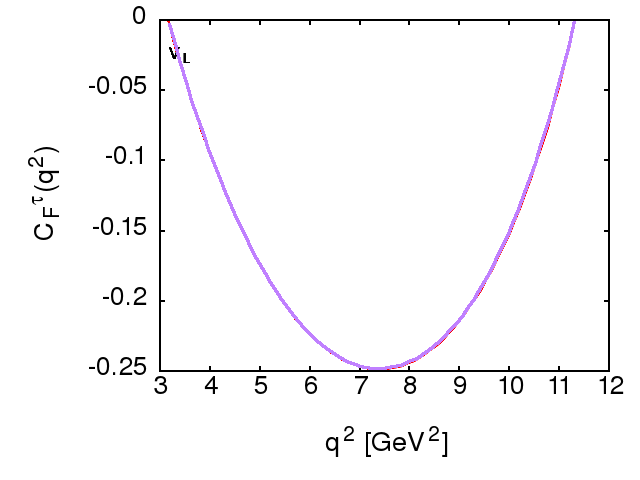}
\includegraphics[width=4.3cm,height=3.3cm]{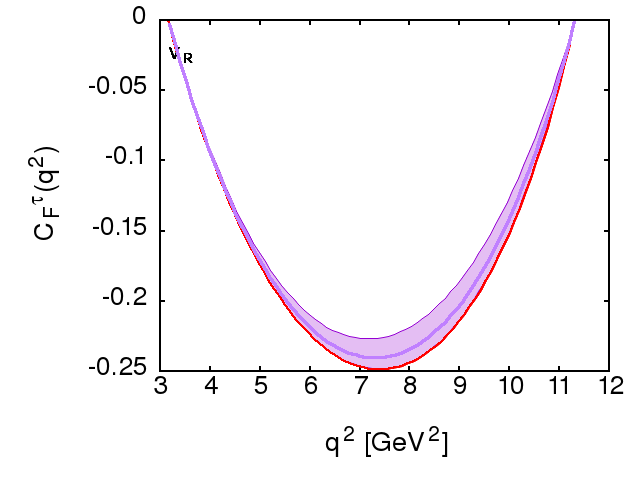}
\includegraphics[width=4.3cm,height=3.3cm]{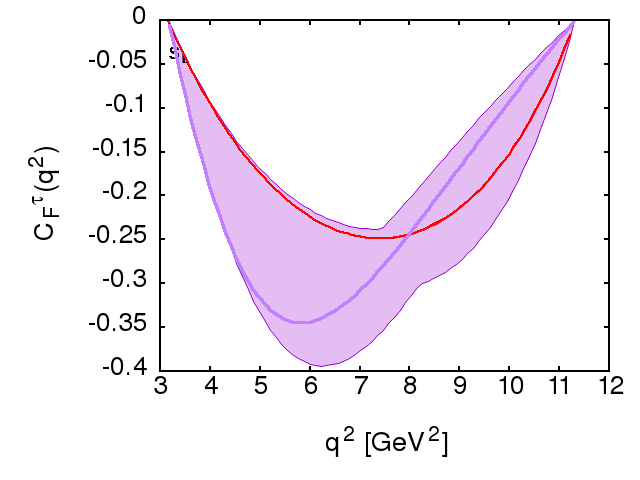}
\includegraphics[width=4.3cm,height=3.3cm]{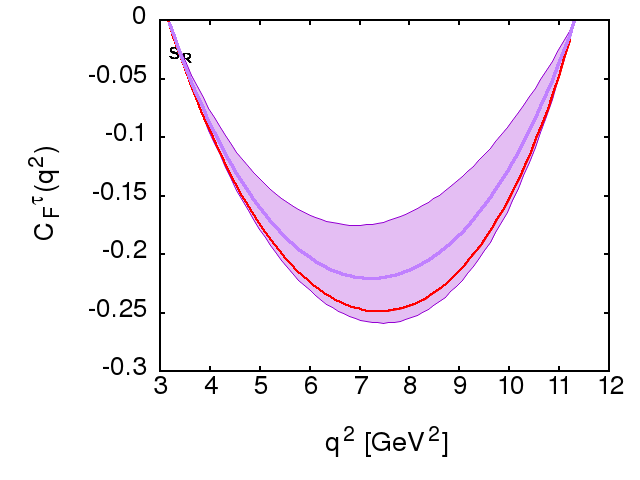}
\caption{The $q^2$ dependency of various observables such as the ratio of branching ratio $R_{\Sigma_c}(q^2)$, 
the total differential decay rate $d\Gamma/dq^2$, the tau polarization fraction 
$P^{\tau}(q^2)$, the forward-backward asymmetry $A_{FB}^{\tau}(q^2)$ and the convexity parameter $C_{F}^{\tau}(q^2)$ for the 
$\Sigma_b \to \Sigma_c \tau \nu$ decay mode in the presence of $V_L$~(first column), $V_R$~(second column), $S_L$~(third column) and 
$S_R$~(fourth column) NP couplings are shown with the purple band, whereas, the SM prediction is shown with red band.
The red solid line represents the SM prediction with the central values of each input parameter and the purple solid line represents
the prediction once the best fit values of the NP couplings are used.}

\label{fignpsig}
\end{figure}

\begin{figure}[htbp]
\centering
\includegraphics[width=4.3cm,height=3.3cm]{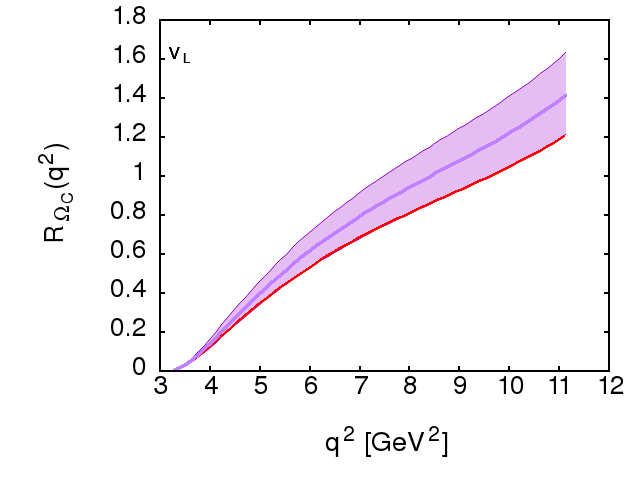}
\includegraphics[width=4.3cm,height=3.3cm]{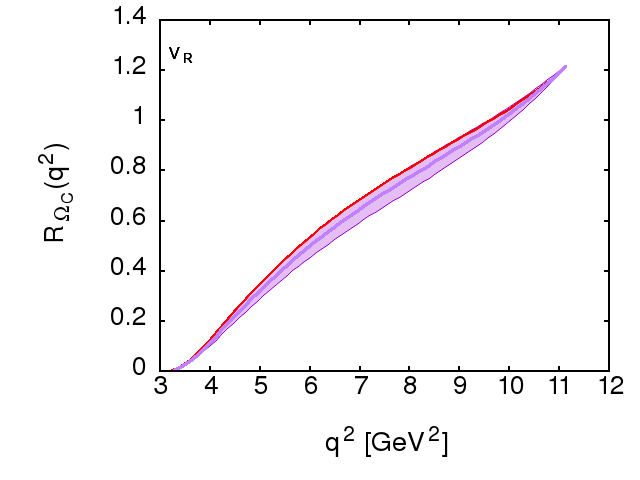}
\includegraphics[width=4.3cm,height=3.3cm]{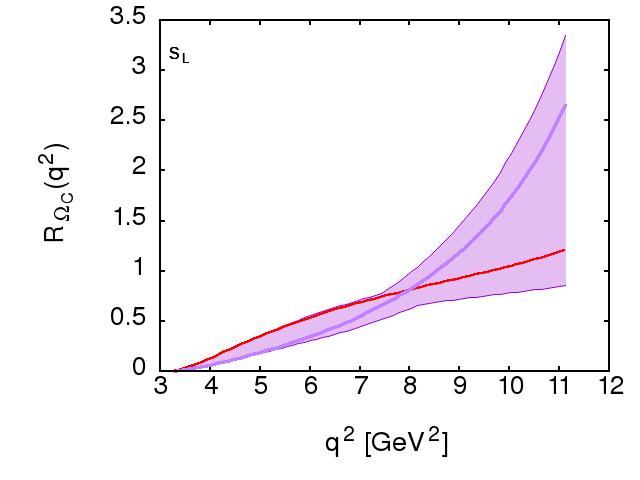}
\includegraphics[width=4.3cm,height=3.3cm]{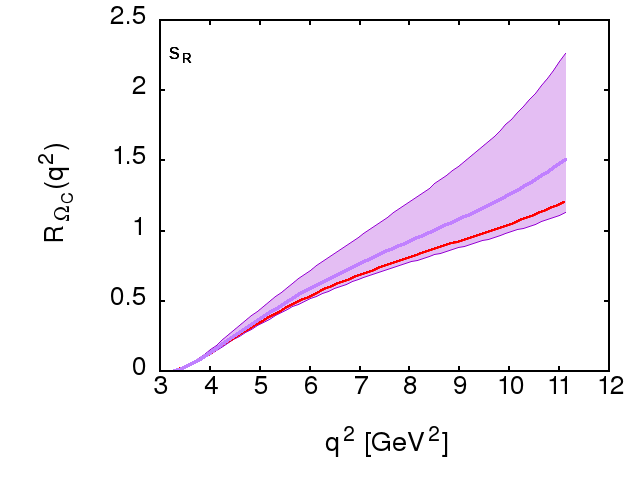}
\includegraphics[width=4.3cm,height=3.3cm]{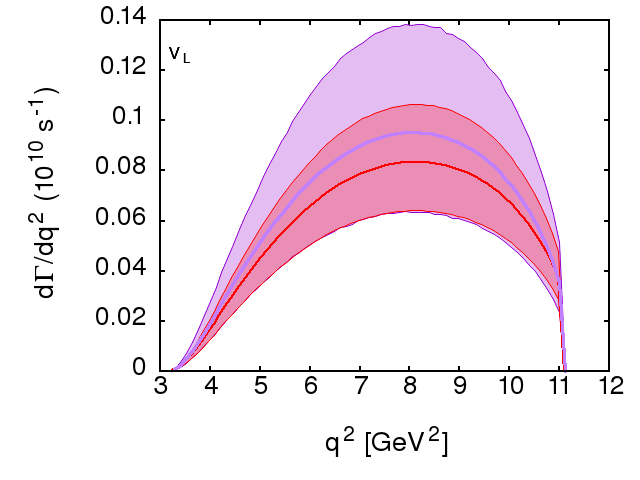}
\includegraphics[width=4.3cm,height=3.3cm]{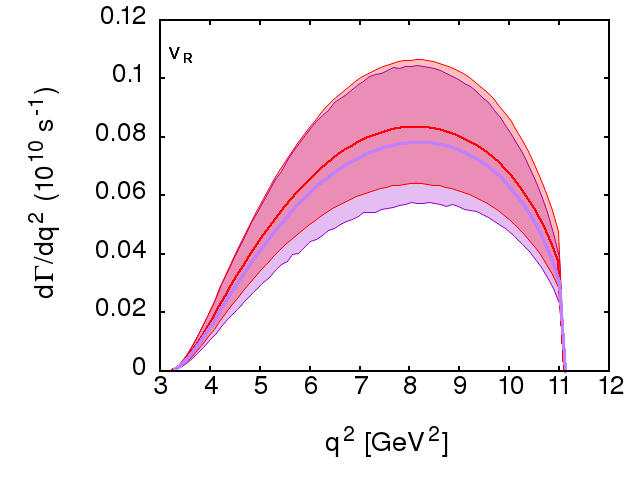}
\includegraphics[width=4.3cm,height=3.3cm]{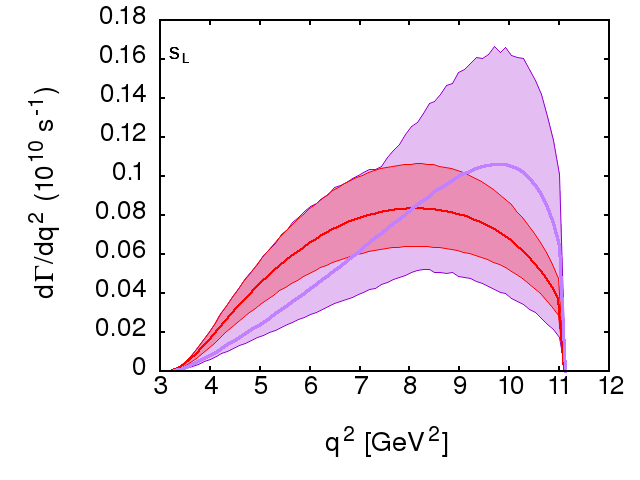}
\includegraphics[width=4.3cm,height=3.3cm]{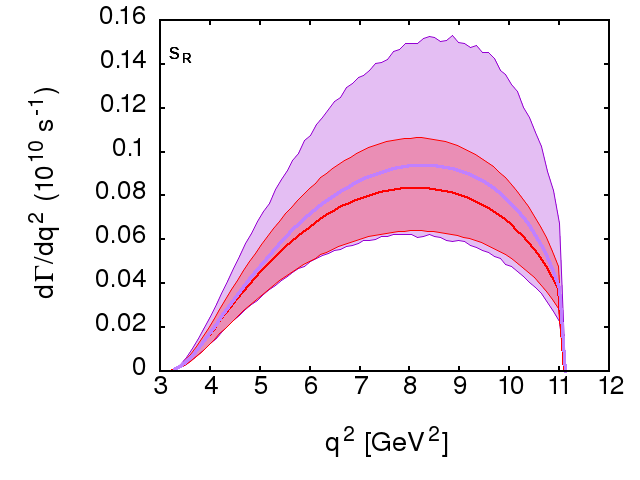}
\includegraphics[width=4.3cm,height=3.3cm]{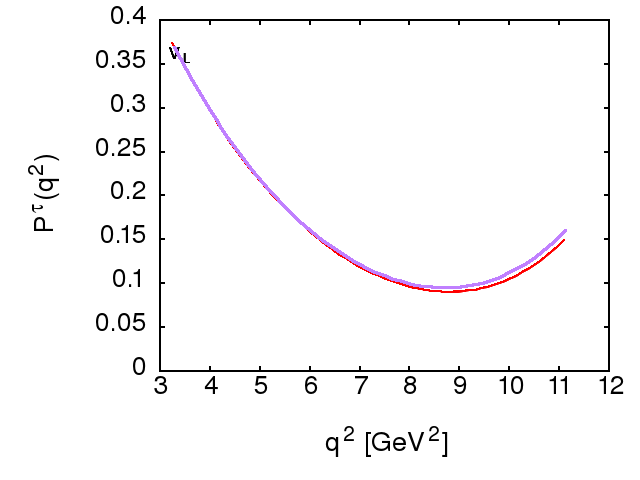}
\includegraphics[width=4.3cm,height=3.3cm]{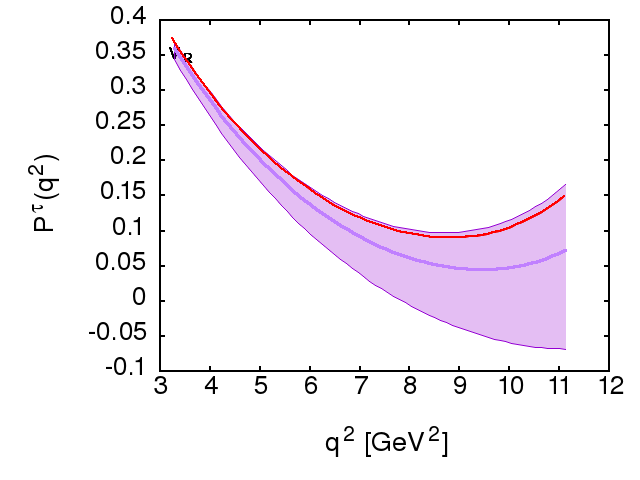}
\includegraphics[width=4.3cm,height=3.3cm]{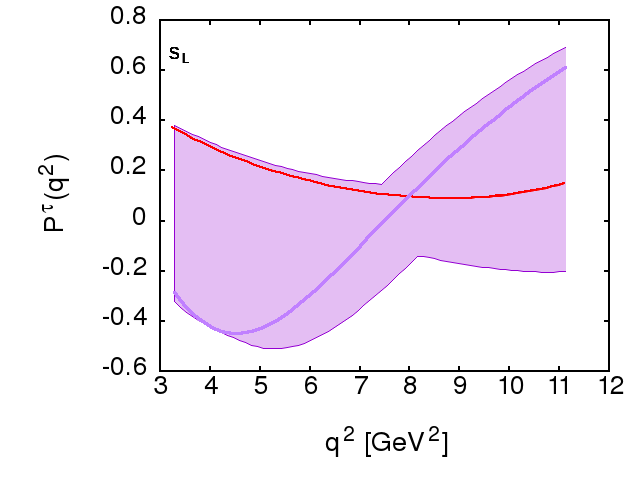}
\includegraphics[width=4.3cm,height=3.3cm]{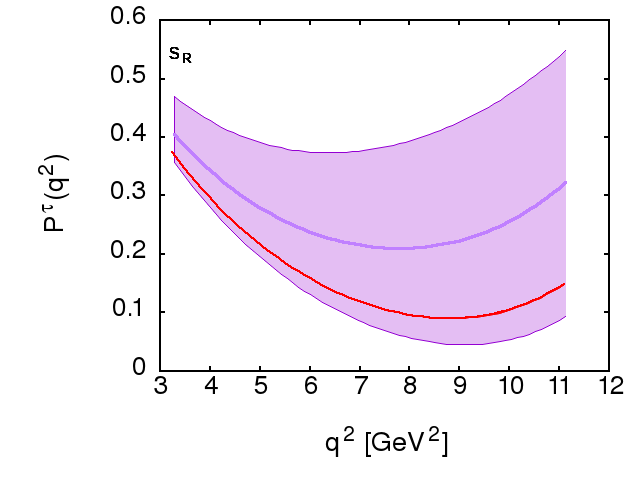}
\includegraphics[width=4.3cm,height=3.3cm]{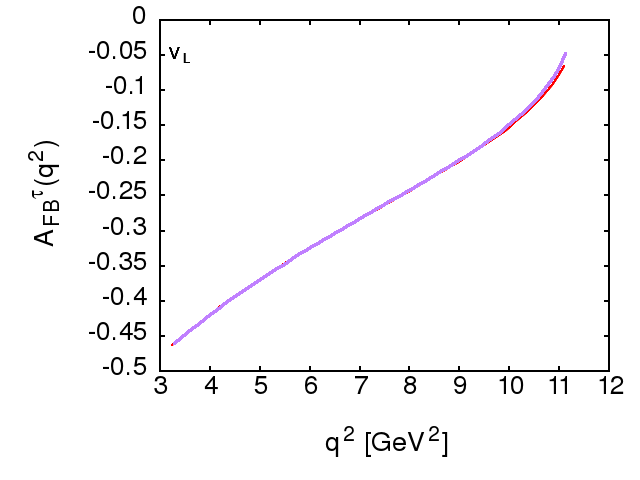}
\includegraphics[width=4.3cm,height=3.3cm]{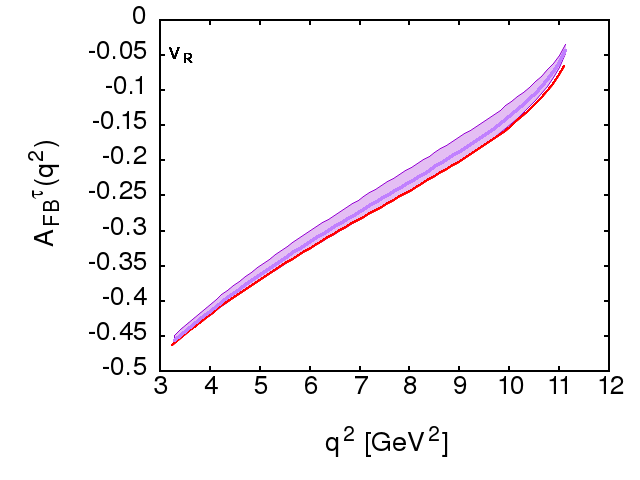}
\includegraphics[width=4.3cm,height=3.3cm]{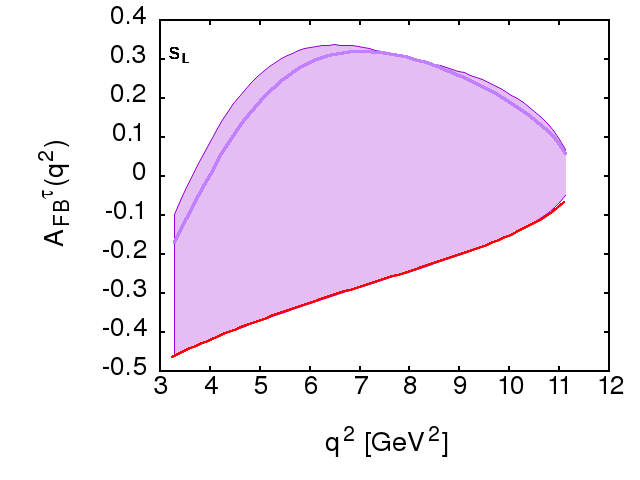}
\includegraphics[width=4.3cm,height=3.3cm]{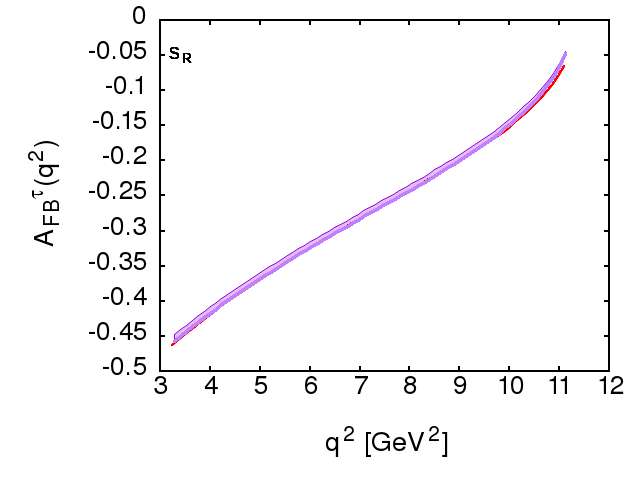}
\includegraphics[width=4.3cm,height=3.3cm]{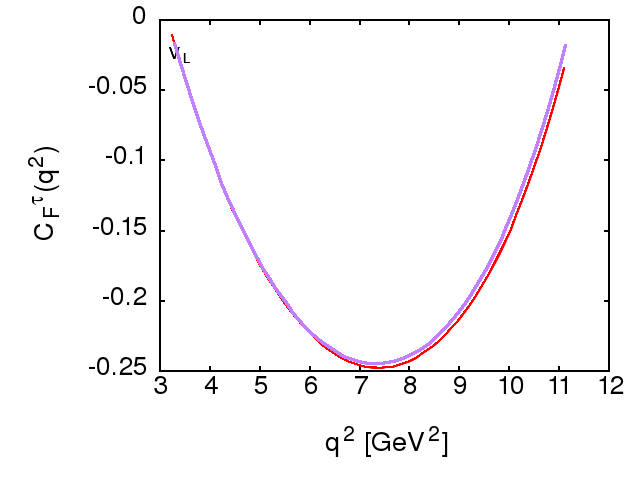}
\includegraphics[width=4.3cm,height=3.3cm]{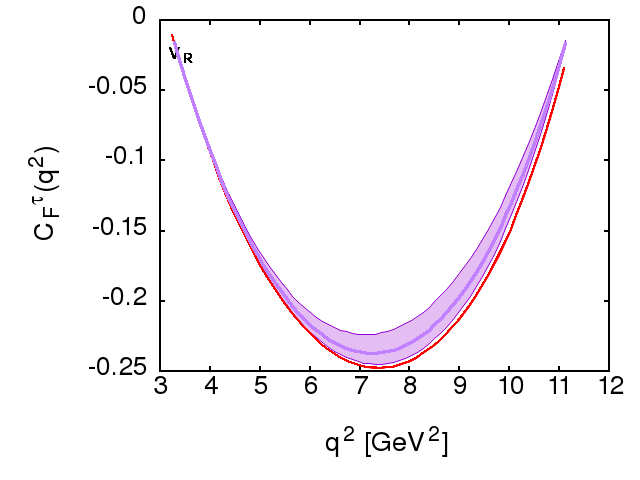}
\includegraphics[width=4.3cm,height=3.3cm]{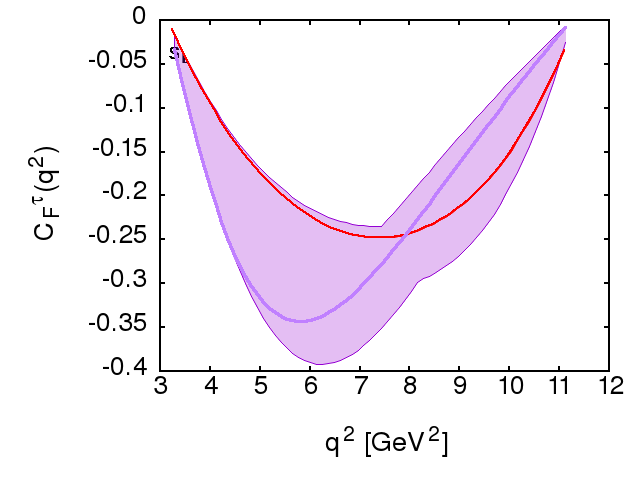}
\includegraphics[width=4.3cm,height=3.3cm]{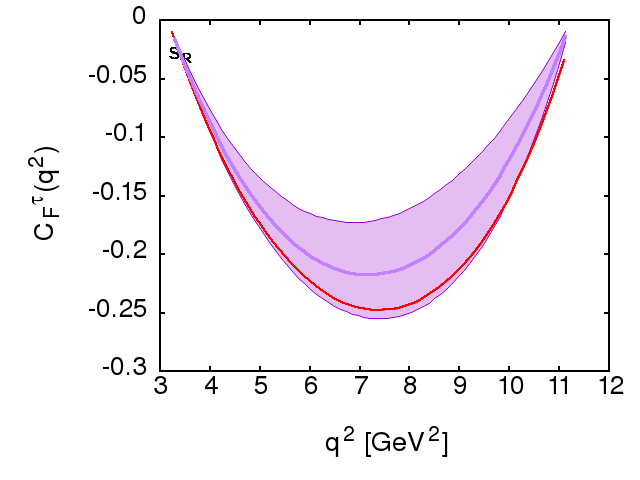}
\caption{The $q^2$ dependency of various observables such as the ratio of branching ratio $R_{\Sigma_c}(q^2)$,
the total differential decay rate $d\Gamma/dq^2$, the tau polarization fraction
$P^{\tau}(q^2)$, the forward-backward asymmetry $A_{FB}^{\tau}(q^2)$ and the convexity parameter $C_{F}^{\tau}(q^2)$ for the 
$\Omega_b \to \Omega_c \tau \nu$ decay mode in the presence of $V_L$~(first column), $V_R$~(second column), $S_L$~(third column) and
$S_R$~(fourth column) NP couplings are shown with the purple band, whereas, the SM prediction is shown with red band.
The red solid line represents the SM prediction with the central values of each input parameter and the purple solid line represents
the prediction once the best fit values of the NP couplings are used.}
\label{fignpomg}
\end{figure}

We also report the $q^2$ dependency of each observable such as the ratio of branching ratio $R_{\Sigma_c}(q^2)$
and $R_{\Omega_c}(q^2)$, the total differential decay rate $d\Gamma/dq^2$, the tau polarization fraction $P^{\tau}(q^2)$, the forward-backward asymmetry
$A_{FB}^{\tau}(q^2)$ and the convexity factor $C_{F}^{\tau}(q^2)$ for both the decay modes in Fig.~\ref{fignpsig} and \ref{fignpomg}.
In each figure we incorporate both SM and NP behavior. The SM and NP are distinguished by red and purple colors respectively.
We represent the SM central curve and the corresponding $1\sigma$ band which we obtain by varying the input parameters (form factors
and $V_{cb}$) within $1\sigma$ with the red color. 
On the other hand, the best fit curve and the band for each NP coupling obtained by imposing the $3\sigma$ constraint coming from the
measured values of $R_D$ and $R_{D^{\ast}}$ are represented with the purple color. 
Our main observations are as follows:

\begin{itemize}
 \item 
The effect of $V_L$ NP coupling is encoded in the vector and axial vector helicity amplitudes only.
In case of $\Sigma_b \to \Sigma_c \tau \nu$ decays, the deviation from the SM prediction due to $V_L$ NP coupling is observed only in 
the ratio of branching ratio $R(q^2)$ and the total differential decay rate $d\Gamma/dq^2$.
All the other observables such as $P^{\tau}(q^2)$, $A_{FB}^{\tau}(q^2)$ and $C_{F}^{\tau}(q^2)$ are SM like. The NP effects get cancelled
in the ratio.
Similar conclusions can be made for the $\Omega_b \to \Omega_c \tau \nu$ decay mode as well.
 
\item Similar to $V_L$, the $V_R$ NP effects are encoded in the vector and the axial vector helicity amplitudes alone.
Deviation in each observable from the SM prediction is observed in this scenario. There is no cancellation of NP effects in 
$P^{\tau}(q^2)$, $A_{FB}^{\tau}(q^2)$ and $C_{F}^{\tau}(q^2)$.
The deviation observed in $d\Gamma/dq^2$, $R(q^2)$, $A_{FB}^{\tau}(q^2)$ and $C_{F}^{\tau}(q^2)$ are less in comparison to the deviation 
observed in the tau polarization fraction $P^{\tau}(q^2)$. Similar conclusions can be made for the $\Omega_b \to \Omega_c \tau \nu$ decay 
mode as well.
 
 \item The scalar NP coupling $S_L$ comes into the decay amplitude through the scalar and pseudoscalar helicity amplitudes.
The deviation observed in this scenario is more pronounced than the deviation observed with $V_L$ and $V_R$ NP couplings.
More interestingly, the SM central curve and the best fit curve due to $S_L$ NP coupling show completely different behavior for
all the observables.
Moreover, there is even a zero crossing in the best fit curve of tau polarization fraction $P^{\tau}(q^2)$ at 
$q^2\approx 7.5$ GeV$^2$ below which $P^{\tau}(q^2)$ takes negative values. 
Similarly, the best fit curve of forward-backward asymmetry $A_{FB}^{\tau}(q^2)$ has a zero crossing around $q^2\approx 3.5$ GeV$^2$. 
However, depending on the value of $S_L$ NP coupling, there may or may not be any zero crossing in $P^{\tau}(q^2)$ and $A_{FB}^{\tau}(q^2)$.
 
 \item Similar to $S_L$, NP effects coming from $S_R$ NP coupling are encoded in the scalar and pseudoscalar helicity amplitudes only. 
Again a significant deviation from the SM prediction is observed, in particular, for $R (q^2)$, $d\Gamma/dq^2$, $P^{\tau}(q^2)$ and 
$C_F^{\tau}(q^2)$. It is, however, worth mentioning that the NP effect in $A_{FB}(q^2)$ is quite negligible in this scenario. Unlike $S_L$,
the shape of each observable remains similar to SM in this scenario.
\end{itemize}

\section{Summary and Conclusion}
\label{summary}

The main objective of this work is to determine the size of the lepton flavor universality violation in the semileptonic decays of
$\Sigma_b$ and $\Omega_b$ heavy baryons. Motivated by the long standing flavor anomalies in $B \to D^{(*)}\, l\, \nu$ decay modes,
we follow a model independent effective field theory approach and study the various physical 
observables within the SM and in the presence of new vector and scalar type NP couplings. We have used the helicity formalism to construct
the angular decay distribution for the $b \to cl\nu$ transitions. We define several observables such as the lepton polarization, lepton side 
forward-backward asymmetry and convexity parameter for the $\Sigma_b \to \Sigma_c l \nu$ and $\Omega_b \to \Omega_c l \nu$ decays.
The numerical results have been presented for both electron mode and tau mode within the SM. We also display the $q^2$ dependant plots within 
SM and within various NP scenarios. 
To find the allowed parameter space, we impose a $3\sigma$ constraint coming from the measured ratio of branching ratios $R_D$ and 
$R_{D^{\ast}}$. We perform our analysis by considering each NP parameter one at time. 
We also perform a naive $\chi^2$ analysis to determine the best fit values of each NP couplings. The corresponding best fit values of 
each observable are also reported. The deviation observed with scalar NP couplings is more pronounced than that with the vector NP couplings.
The deviation observed in case of $S_L$ NP coupling is quite distinct from all other NP couplings. In future, this may help to identify 
the exact nature of NP.

Unlike $B$ meson decays which are
rigorously studied both theoretically and experimentally over the decade, the baryonic decay modes which undergo similar quark level 
transitions are less explored. Study of these decay modes are useful for two reasons. First, it can provide us complementary 
information regarding NP in various $B$ meson decays and also can be useful in determining the value of the CKM matrix element $|V_{cb}|$. 
Secondly, study of these decay modes both theoretically and experimentally can act as a useful ingredient in maximizing future sensitivity to
NP.

\section*{Acknowledgment}
We are grateful to D. Ebert, R. N. Faustov and V. O. Galkin for providing us the Isgur-Wise functions $\zeta_1(w)$ and $\zeta_2(w)$
in the whole kinematic range used for computing $\Sigma_b \to \Sigma_c$ and $\Omega_b \to \Omega_c$ transition form factors.

\end{document}